\documentclass[unnumsec,webpdf,contemporary,large]{oup-authoring-template}%

\usepackage{graphicx}
\usepackage{xcolor}
\usepackage{hyperref}
\usepackage{url}
\usepackage{multirow}
\usepackage{caption}
\usepackage{subcaption}
\usepackage{booktabs}
\usepackage{multicol}
\usepackage{colortbl}

\newcommand\blfootnote[1]{%
  \begingroup
  \renewcommand\thefootnote{}\footnote{#1}%
  \addtocounter{footnote}{-1}%
  \endgroup
}

\usepackage{enumitem}
\begin{document}

\journaltitle{Bioinformatics}
\DOI{DOI HERE}
\copyrightyear{2023}
\pubyear{2023}
\access{Advance Access Publication Date: Day Month Year}
\appnotes{Paper}

\firstpage{1}

%\subtitle{Subject Section}

\title[DNA - Embed Sequence Align]{Embed-Search-Align: DNA Sequence Alignment using Transformer models}

% \author[1,$\ast$]{First Author}
% \author[2]{Second Author}
% \author[3]{Third Author}
% \author[3]{Fourth Author}
% \author[4]{Fifth Author\ORCID{0000-0000-0000-0000}}

% \authormark{Author Name et al.}

% \address[1]{\orgdiv{Department}, \orgname{Organization}, \orgaddress{\street{Street}, \postcode{Postcode}, \state{State}, \country{Country}}}
% \address[2]{\orgdiv{Department}, \orgname{Organization}, \orgaddress{\street{Street}, \postcode{Postcode}, \state{State}, \country{Country}}}
% \address[3]{\orgdiv{Department}, \orgname{Organization}, \orgaddress{\street{Street}, \postcode{Postcode}, \state{State}, \country{Country}}}
% \address[4]{\orgdiv{Department}, \orgname{Organization}, \orgaddress{\street{Street}, \postcode{Postcode}, \state{State}, \country{Country}}}

\author[1,*,\textdagger]{Pavan Holur}
\author[2,3,\textdagger]{K. C. Enevoldsen}
\author[1]{Shreyas Rajesh}
\author[4]{Lajoyce Mboning}
\author[5]{Thalia Georgiou}
\author[4]{Louis-S. Bouchard}
\author[6]{Matteo Pellegrini}
\author[1, \textdagger]{Vwani Roychowdhury}

\address[1]{\orgdiv{Department of Electrical and Computer Engineering}, \orgname{UCLA}}
\address[2]{\orgdiv{Center for Humanities Computing}, \orgname{Aarhus University}}
\address[3]{\orgdiv{Center for Quantitative Genetics and Genomics}, \orgname{Aarhus University}}
\address[4]{\orgdiv{Department of Chemistry and Biochemistry}, \orgname{UCLA}}
\address[5]{\orgdiv{Department of Biochemistry, Biophysics, and Structural Biology (MBIDP)}, \orgname{UCLA}}
\address[6]{\orgdiv{Molecular, Cell, and Developmental Biology}, \orgname{UCLA}}

% \texttt{\{pholur,lajoycemboning,thaliageorgiou,matteop,vwani\}@ucla.edu,}\\ \texttt{kenneth.enevoldsen@cas.au.dk, louis.bouchard@gmail.com} 

\corresp[$\ast$]{Corresponding author. \href{pholur@g.ucla.edu}{pholur@g.ucla.edu}}

\received{Date}{0}{Year}
\revised{Date}{0}{Year}
\accepted{Date}{0}{Year}

%\editor{Associate Editor: Name}

%\abstract{
%\textbf{Motivation:} .\\
%\textbf{Results:} .\\
%\textbf{Availability:} .\\
%\textbf{Contact:} \href{name@email.com}{name@email.com}\\
%\textbf{Supplementary information:} Supplementary data are available at \textit{Journal Name}
%online.}

\abstract{DNA sequence alignment, an important genomic task, involves assigning short DNA reads to the most probable locations on an extensive reference genome. 
%It is used in various genomic analyses, including variant calling, transcriptomics, and epigenomics. 
%This process is crucial for various genomic analyses, including variant calling, transcriptomics, and epigenomics. 
Conventional methods tackle this challenge in two steps: genome indexing followed by efficient search to locate likely positions for given reads. Building on the success of Large Language Models (LLM) in encoding text into embeddings, where the
distance metric captures semantic similarity, recent efforts have encoded DNA sequences into vectors using Transformers and have shown promising results in tasks involving classification of short DNA sequences. 
%Building on the success of Large Language Models (LLM) in encoding text into embeddings, where the distance metric captures semantic similarity, recent efforts have explored whether one can generate similar numerical representations for DNA sequences. Such models have shown early promise in tasks involving classification of short DNA sequences.  
%such as the detection of coding- vs non-coding regions, as well as the identification of enhancer and promoter sequences. 
Performance at sequence classification tasks does not, however, guarantee \textit{sequence alignment}, where it is necessary to conduct a genome-wide search to align every read successfully, a  \textit{significantly longer-range task by comparison}. We bridge this gap by developing a ``\textbf{E}mbed-\textbf{S}earch-\textbf{A}lign'' (ESA) framework, where a novel Reference-Free DNA Embedding (\textit{RDE}) Transformer model generates vector embeddings of reads and fragments of the reference in a shared vector space; read-fragment distance metric is then used as a surrogate for sequence similarity. %representations of reads and fragments of the reference, which are projected into a shared vector space where the read-fragment distance is used as a surrogate for alignment. 
ESA introduces: (1) Contrastive loss for self-supervised training of DNA sequence representations, facilitating rich reference-free, sequence-level embeddings, and (2) a DNA vector store to enable search across fragments on a global scale. RDE is 99\% accurate when aligning 250-length reads onto a human reference genome of 3 gigabases (single-haploid), \textcolor{black}{rivaling conventional algorithmic sequence alignment methods such as \textit{Bowtie} and \textit{BWA-Mem}}. RDE far exceeds the performance of $6$ recent DNA-Transformer model baselines \textcolor{black}{such as \textit{Nucleotide Transformer, Hyena-DNA}}, and shows task transfer across chromosomes and species.}

% \boxedtext{
% \begin{itemize}
% \item Key boxed text here.
% \item Key boxed text here.
% \item Key boxed text here.
% \end{itemize}}

\maketitle

\section{1. Introduction}

\blfootnote{\textdagger \textit{Equal Contribution}}

Sequence alignment is a central problem in the analysis of sequence data.  It is used in various genomic analyses, including variant calling, transcriptomics, and epigenomics. Many DNA sequencers generate short reads that are only a couple of hundred bases long.  In order to interpret this data, a typical first step is to align the reads to a genome. Genomes come in many sizes, and commonly studied genomes range from millions of bases (e.g., bacteria) to billions of bases (e.g., mammals, plants) in length. The resulting task of aligning short reads to large-size genomes is  computationally challenging---akin to finding a needle in a haystack---and many decades of work have led to optimized approaches that can perform these tasks with great efficiency. 
%Moreover, most experiments will generate millions of reads.  
%However, many limitations remain with existing genome alignment methods.  One limitation is that a reference genome usually is an approximation of the sequence from a single individual, and does not capture the variability of genomes across populations.  Another, is that sequence alignment algorithms may not correctly model sequencing errors, and thus may introduce mistakes when trying to score the best match for a sequence against a genome.  We asked whether some of these limitations could potentially be overcome if 
Since genomes are  sequences  where the alphabet consists of only a few symbols ($\{A,T,G,C\}$) they can be considered as Limited Alphabet Languages (LAL). We ask whether one can design a new paradigm to align reads to genomes by exploiting the Transformer architectures~\cite{vaswani_attention_2017} \textcolor{black}{adopted for recent language modeling efforts}. The applications of the Transformer model---and, more generally ``Large Language Models'' (LLM)---in Bioinformatics applications are still in their infancy, yet hold substantial promise: Transformer models have demonstrated an unprecedented ability to construct powerful numerical representations of sequential data~\cite{vit, wav2vec, gpt3}.

We establish a foundation model~\cite{foundation}--Reference-free DNA Embedding (RDE)---tailored for embedding DNA sequences.  The model $h_\Theta$ (i.e., parameterized by weights $\Theta$)  maps any genome subsequence $F$ into an embedding $h_\Theta(F) \in \mathcal{R}^n$ (e.g., $n=768)$, such that if $F_1$ and $F_2$ are similar subsequences of differing lengths---for example, $F_1$ is any noisy subsequence of $F_2$ or $F_1$ and $F_2$ have significant overlaps---then the $d(h_\Theta(F_1),  h_\Theta(F_2))$ is small; where $d(\cdot)$ is a distance metric in the embedding space, such as the cosine distance between two vectors. Another requirement of such RDE models is that they should generate embeddings that are reference free, i.e. once trained, it creates semantically-aware embedding (i.e. sequences that are close in edit distance are mapped close to each other) of any given genome subsequences \textit{irrespective of the reference genome from which it is sampled}. Given such a model, alignment of reads can be achieved by a near-neighbor search in the embedding space: only the fragments of genome with embeddings nearest to the embedding of the given read need to be considered. Thus, a global search in a giga-bases long sequence is reduced to a local search in a vector space.

\begin{figure*}[t]
    \centering
    \includegraphics[width=1.0\textwidth]{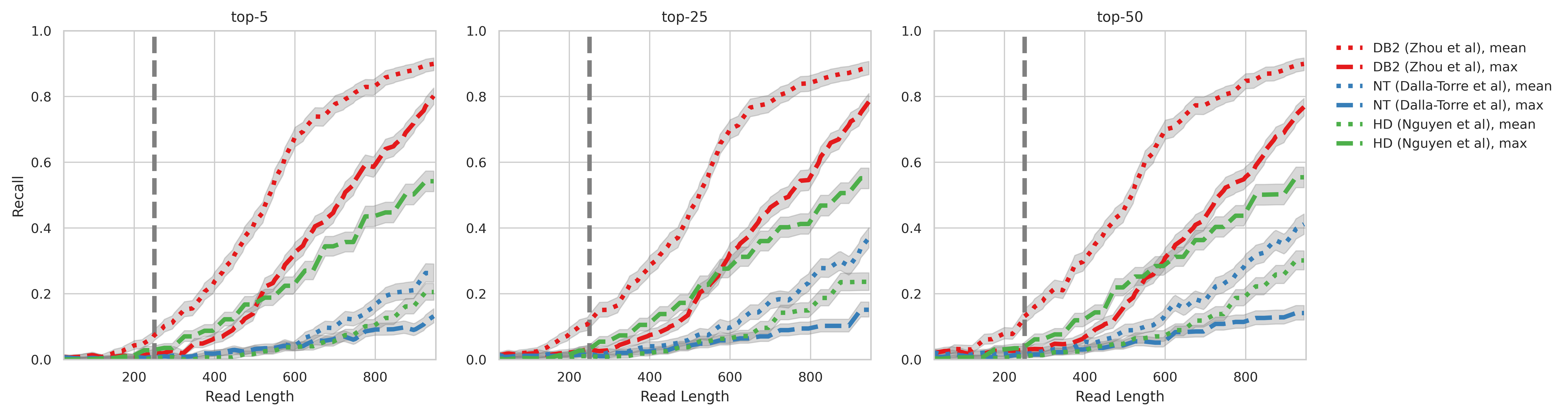}
    \caption{\textbf{Alignment Recall of Transformer-DNA Baselines by Read Length:} Existing Transformer-DNA models were adapted for sequence alignment using mean-/max-pooling. Their performance, measured by recall (top-$K$) over $40K$ reads of varying lengths across the human genome (3gb - single-haploid), is shown. Trendlines represent each baseline, with error bars (Clopper-Pearson Interval~\citep{cps} @ 95\%) in grey. The vertical line at $x=250$ marks a typical read length. Overall, these baselines show suboptimal performance. For more details, see Sec.~\ref{sec:alignpure}.}
    \label{fig:puretests}
    \vspace{-1em}
\end{figure*}

% Computational simulators have been developed to generate synthetic reads that have properties of real reads. These simulators mimic the read quality and characteristics produced by actual sequencing machines, thus providing a scalable means for validating new alignment approaches~\citep{huang_art_2012}.

\begin{figure*}[t]
     \centering
     \begin{subfigure}[t]{0.48\textwidth}
         \centering
         \includegraphics[width=0.6\textwidth]{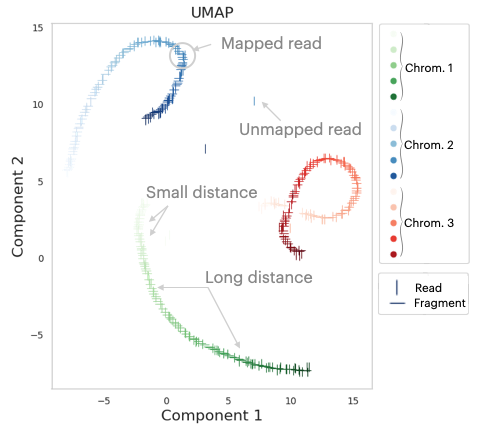}
         \caption{Three $20,000$-long nucleotide sequences were selected, one each from Chr. 1, 2, and 3. Each sequence is broken up into $100$ \textit{consecutive} reference fragments, each of length $1000$. Consecutive fragments have an overlap of $800$ bases (stride $200$). 
         Additionally, reads of length $Q \sim \mathcal{U}[100,250]$ are sampled from within each fragment. The colors correspond to the respective chromosomes and the position of a particular fragment along the 20K-nucleotide sequence is coded by its color intensity.}
         \label{fig:emb3}
     \end{subfigure}
     \hfill
\hfill
\begin{subfigure}[t]{0.48\textwidth}
\centering
\includegraphics[width=0.6\textwidth]{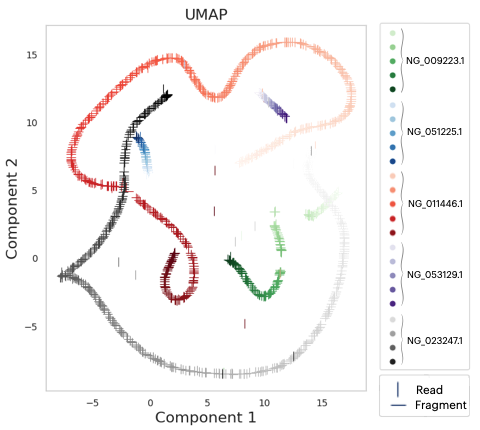}
\caption{Five gene regions (listed above in the inset) of different sizes are selected from the reference genome. Each gene is broken up into consecutive fragments of length $1000$ with a stride of $200$ (similar to subfigure (a)). Additionally, reads of length $Q \sim \mathcal{U}[100,250]$ are sampled from within each fragment. The colors correspond to genes and the position of a particular fragment along a specific gene is coded by its color intensity.}
\label{fig:emb1}
\end{subfigure}

\caption{\textbf{Illustrating RDE's Preservation of Sequence Locality in Embedding Space:} The reference genome $\mathcal{R}$ is divided into fragments $\mathcal{F}_i$, each represented by an embedding $h(\mathcal{F}_i)$. For effective sequence alignment, specific structures are expected in the embedding space: (1) Overlapping fragments $\mathcal{F}_i$ and $\mathcal{F}_j$ should have proximate embeddings; (2) Consecutive fragments forming a long sequence should correspond to a distinct manifold in the embedding space. Subfigures (a) and (b) display this emergent geometry, as visualized using embeddings produced by RDE. In both subfigures, the fragment and read representations are jointly visualized in a 2D low-dimensional space constructed by applying  UMAP~\cite{umap} to the original embeddings. Vertical marker lines ($|$) refer to fragments, ``--'' markers correspond to reads. We observe that the consecutive fragments belonging to the same nucleotide (subfigure (a)) or gene (subfigure (b)) sequence constitute an order-preserving 1D manifold. Additionally, \textit{almost all of the reads} are close to their corresponding fragments, making this space viable for the task of Sequence Alignment: by searching for fragments in the neighborhood of an external read represented in the embedding space, one is likely to retrieve a fragment most responsible for the read.}
\label{fig:eyecandy}
%\vspace{-1em}
\end{figure*}

\subsection{1.1 Transformer models: Written Language to DNA Sequence Alignment}

DNA sequences share remarkable similarities with written language, offering a compelling avenue for the application of Transformer models. Like written language, these are sequences generated by a small alphabet of nucleotides $\{A, T, G, C\}$. Traditional DNA modeling efforts have already accommodated mature encoding and hashing techniques initially developed for written language -- such as Suffix trees/arrays and Huffman coding~\citep{huffman_method_1952,manber_suffix_1993} -- to successfully parse and compress DNA sequences. Alignment methods such as MinHash and MashMap~\cite{mashmap} have incorporated \textit{Locality Sensitive} Hashing (LSH)~\cite{lsh} to construct fast, approximate representations of DNA sequences that facilitate the rapid matching of millions of reads onto the reference genome. 
%Furthermore, just as written language contains repeated subsequences (words, phrases) to represent real-world objects, DNA sequences similarly possess repeating ``words'' and groupings of such words into a ``sentence'' representing, for example, subsets of genes.

Within the last few years, \color{black} several Transformer-based models have been developed for DNA sequence analysis. Notably, DNABERT-2~\citep{ji_dnabert_2021, dna2bert}, Nucleotide Transformer~\citep{nucleotide}, GenSLM~\citep{zvyagin_genslms_2022}, HyenaDNA~\citep{nguyen_hyenadna_2023} and GENA-LM~\citep{fishman_gena-lm_2023} have been designed to discern relationships between short genetic fragments and their functions. Specifically, Nucleotide Transformer representations have shown utility in classifying key genomic features such as enhancer regions and promoter sequences. Similarly, GENA-LM has proven effective in identifying enhancers and Poly-adenylation sites in Drosophila. In parallel, DNABERT-2 representations have also been found to cluster in the representation space according to certain types of genetic function. 

These models, \textcolor{black}{trained on classification tasks}, generate sequence embeddings such that their pairwise distances correspond to class separation. \textcolor{black}{As a result,} pairs of sequences with very large edit distances between them \textcolor{black}{are mapped to} numerical representations that are close by. However, in tasks such as Sequence Alignment \textcolor{black}{the objective is quite different:} \textit{the pairwise representation distance has to closely match the sequence edit distance}.  A natural question arises: \textit{Can these Transformer architectures be readily applied to the task of Sequence Alignment?} We delineate the associated challenges as follows:

\begin{itemize}[leftmargin=2.4em, topsep=2pt,itemsep=0pt,partopsep=0pt, parsep=0pt]
\item[\textbf{[L1]}] ~ \textbf{Two-Stage Training:} DNA-based Transformer models typically undergo pretraining via a \textit{Next Token/Masked Token Prediction} framework, a method originally developed for natural language tasks. To form sequence-level representations, these models often employ pooling techniques that aggregate token-level features into a single feature vector. This approach, however, is sometimes critiqued for yielding suboptimal aggregate features ~\citep{reimers_sentence-bert_2019}.
  
\item[\textbf{[L2]}]  ~ \textbf{Computation Cost:} The computational requirements for Transformer models grow quadratically with the length of the input sequence. This is particularly challenging for sequence alignment tasks that necessitate scanning entire genomic reference sequences.
\end{itemize}

Figure~\ref{fig:puretests} shows the sequence alignment performance  (recall) of several Transformer-DNA models. The testing protocols are elaborated in Sec.~\ref{sec:alignpure}. Notably, these models exhibit subpar recall performance when aligning typical read lengths of 250.

\section{2. Our Contributions}

In this paper, we argue that both limitations \textbf{L1}, \textbf{L2} of Transformer-DNA models can be mitigated by formulating sequence alignment as a vector search-and-retrieval task. Our approach is twofold: (A) We introduce a sequence encoder \textit{Reference-Free DNA Embedding} (RDE), trained through self-supervision, to embed DNA subsequences to vectors.  (B) Next, we formulate a framework, \textit{Embed-Search-Align} (ESA), that maps reads to  reference genome sequence. We leverage a specialized data structure, termed a \textit{DNA vector store}, that enables efficient storage and search in an embedding space\footnote{Codebase available here: \url{https://anonymous.4open.science/r/dna2vec-7E4E/}}: the entire reference genome is sharded into equal-length overlapping fragments whose embeddings are then uploaded to the vector store. For each read, top-K closest fragments can then be searched efficiently in time that scales only logarithmically in the number of fragments \cite{hnsw}. These strategies have been explored in NLP: (A) Sequence-to-embedding training using contrastive loss has shown improved performance---over explicit pooling methods---at abstractive semantic tasks such as evidence retrieval~\cite{lewis} and semantic text similarity~\citep{gao_simcse_2021, chen_simple_2020}.  (B) Specialized data structures, such as ``vector stores'' or ``vector databases'' like FAISS~\citep{johnson_billion-scale_2019} and \href{www.pinecone.io}{\textit{Pinecone}}, use advanced indexing and retrieval algorithms for scalable numerical representation search.

\subsection{2.1 Task Definition}

The simplest sequence alignment task applies to single-end\footnote{A DNA fragment is ligated to an adapter and then sequenced from one end only.} reads, where a sequencer generates a read of length $Q$
\begin{align}
r := (\tilde{b}_1, \tilde{b}_{2}, \dots, \tilde{b}_{Q}), 
\label{eq:read}
\end{align}
where $\tilde{b}_i \in \{A, T, G, C\}$. In practice, these reads come from individual genomes that do not necessarily match the reference and may contain mutations due to base insertions, deletions, and substitutions. Thus, it is assumed that this read is a noisy substring taken from a reference genome sequence $\mathcal{R} := (b_1, b_2, \dots, b_N), \, b_i \in \{A, T, G, C\}$ and $N\gg Q$; for example, for the single-haploid human genome~\citep{chm_consort}, $N \approx 3$ gigabases (gb), and $Q\approx 250$ though reads of much longer lengths are becoming increasingly affordable and accurate~\cite{giab}. The alignment task is to find a substring in $\mathcal{R}$
\begin{align}
\tilde{r} := (b_q, b_{q+1}, \dots, b_{q+Q}), \,\, 1 \le q \le N-Q,
\label{eq:readn}
\end{align}
such that the edit distance $d(r,\tilde{r})$---computed using the Smith-Waterman (SW) algorithm---is minimized. The primary objective is to identify the most probable location, $q$, of this read within the reference genome. 

We formulate the problem of Sequence Alignment as minimizing a sequence alignment function, $\texttt{SA}$, applied to a read $r$ and a reference sequence $\mathcal{R}$ as
\begin{align}
v^* = \min_{q} \texttt{SA}(r, \mathcal{R})
\end{align}
where $q \in \{1, 2, 3, \dots\}$ is a candidate reference starting position and $v^*$ is the optimal alignment score. Lower scores indicate better alignments. This optimization exhibits the following property:

\textbf{Sharding for sequence alignment:} For a read segment $r$ of length $Q$ and reference $\mathcal{R}$ of length $N$, the complexity of  $\texttt{SA}(r, \mathcal{R})$ scales as $\mathcal{O}(NQ)$, which is expensive. % when $N \rightarrow Q$.

Ideally, we would like to use $\mathcal{R}$ to get a set of fragments $\{\mathcal{F}_1, \mathcal{F}_2, \dots, \mathcal{F}_K\}$ that are a subset of the original reference. Then:

\begin{align}
v^* \approx \min_{ \mathcal{F}_j \in \{\mathcal{F}_1, \mathcal{F}_2, \dots, \mathcal{F}_K\}} \texttt{SA}\left( r, \mathcal{F}_j \right).
\label{eq:main_opt} 
\end{align}
Here, each $\mathcal{F}_j$ is a fragment of $\mathcal{R}$ (i.e., $\mathcal{F}_j \subset \mathcal{R}$), and $K$ is the number of these sub-tasks. This approximation is effective under the conditions: 
\begin{itemize}[leftmargin=2.4em, topsep=0pt,itemsep=0pt,partopsep=0pt, parsep=0pt]
\item[(1)] ~ Fragment $\mathcal{F}_j$ lengths are on the order of the read length ($Q$), and not the length of the longer reference $\mathcal{R}$;
\item[(2)] ~ Since a read can originate from any position along the reference, there must be sufficient fragments $\mathcal{F}_i$ to cover $\mathcal{R}$, i.e. $\cup \mathcal{F}_j = \mathcal{R}$. From within this set of fragments, we hope to retrieve a subset of fragments containing the optimal fragment for the read;
\item[(3)] ~ While retrieving a subset of fragments of size $K$, we require $K$ to be significantly smaller than $\frac{N}{Q}$. If $\frac{N}{Q}$, then this amounts to scanning the whole reference.
\end{itemize}

\begin{figure*}[t]
    \centering
    \includegraphics[width=0.8\textwidth]{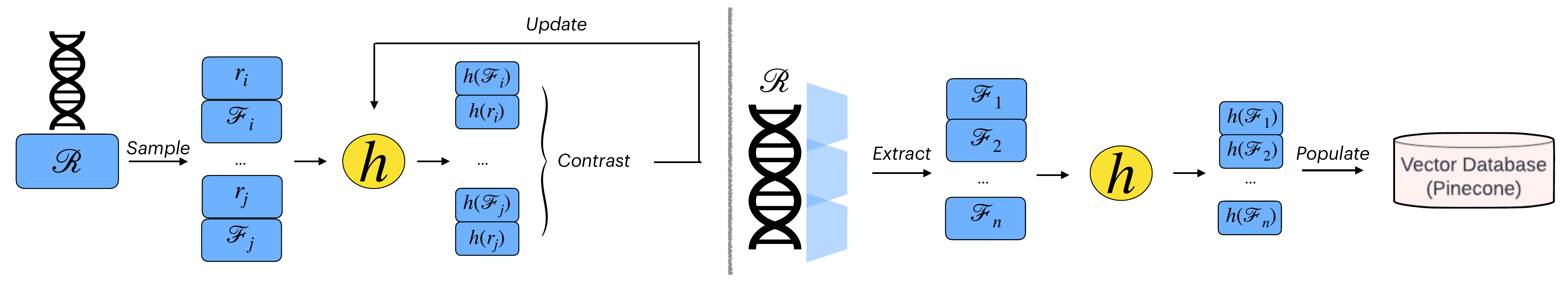}
    \caption{
    \textbf{System Overview [A] - Training Encoder and Populating Vector Store:} Reference genome fragments $\mathcal{F}_i$ and within them, randomly sampled pure reads $r_i$ (positive pairs) are numerically represented via shared encoder $h$. Encoder training follows a contrastive approach as per \eqref{eq:loss}. After training, the genome is segmented into overlapping fragments, encoded, and uploaded into the vector store.}
    \label{fig:up}
\end{figure*}

Conditions (1) and (2) imply that fragments should be short and numerous enough to cover the reference genome. Condition (3) restricts the number of retrieved reference fragments per
read---that we deem to be most likely to contain $r$---to a small value $K$. Analogous methods have shown efficacy in text-based Search-and-Retrieval tasks ~\citep{peng2023, dai2022} on Open-Domain Question-Answering, Ranking among other tasks. Subsequent sections describe a parallel framework for retrieving reference fragments given a read. The pipeline is shown in  Figs.~\ref{fig:up} and \ref{fig:down}.

\subsection{2.2 RDE: Designing effective sequence representations}

An optimal sequence encoder model $h$ is such that the corresponding embeddings of any read $r$ and reference fragment $\mathcal{F}$---$h(r), h(\mathcal{F})$ respectively---obey the following constraints over a pre-determined distance metric $d$:
\begin{align}
d \{ h(r_j), h(\mathcal{F}_i) \} \ge d \{ h(r_j), h(\mathcal{F}_j)\}, \quad i \neq j
\end{align}

%\begin{align}
%\text{dist} \left\{ h(r), h(\mathcal{F}_i) \right\} > \text{dist} \left\{ h(r), h(\mathcal{F}_*)\right\}, \quad
%\text{if } r \subset \mathcal{F}_*, r \not\subset \mathcal{F}_i, \forall \mathcal{F}_i \neq \mathcal{F}_*.
%\end{align}

 Subscript $j$ serves to indicate that the read $r_j$ is \textit{correctly aligned to} a fragment reference, $\mathcal{F}_j$, (\textit{that we call a positive \{read, fragment\} pair}). Given another arbitrary fragment $\mathcal{F}_i$ to which $r_j$ has poor alignment, $\{r_j, \mathcal{F}_i\}$ is a negative \{read, fragment\} pair. Observe that these inequalities constitute the only requirements for the encoder. \color{black} As long as the \textit{neighborhood} of $r_j$ in the representation space \textit{contains} the representation for $\mathcal{F}_j$, it will be recovered in the nearest neighbors (top-$K$ set) and alignment will succeed. Equality is observed when $r_j$ is a repeat sequence matched equally well to more than one fragment. This motivates using self-supervision~\citep{hadsell_dimensionality_2006, chen_simple_2020, gao_simcse_2021} where we are only concerned about the relative distances between positive and negative (read, reference fragment) pairs.

\subsection{2.3 RDE: Self-supervision and contrastive loss}

A popular choice for sequence learning using self-supervision involves a contrastive loss setup described by ~\cite{chen_simple_2020} and \cite{gao_simcse_2021}: i.e. for a read $r$ aligned to reference fragment $\mathcal{F}_j$, the loss $l_{r}$ simultaneously minimizes the distance of $h(r)$ to $h(\mathcal{F}_j)$ and maximizes the distance to a batch of random fragments of size $B-1$:
\begin{equation}
l_{r} = - \log \frac{e^{-d(h(r), h(\mathcal{F}_j)) /\tau}}{
e^{-d(h(r), h(\mathcal{F}_j)) /\tau}        
+ \sum^{B-1}_{i=1}e^{-d(h(r), h(\mathcal{F}_i))/\tau} 
}. \label{eq:loss}
\end{equation}

Here $\tau$ is a tuneable temperature parameter. To stabilize the training procedure and reach a non-trivial solution, the encoder applies different dropout masks to the reads and fragments similar to the method described in prior work~\cite{chen_simple_2020}. Similar setups have been shown to work in written language applications, most notably in Sentence Transformers~\citep{reimers_sentence-bert_2019,gao_simcse_2021,muennighoff_mteb_2023}, which continue to be a strong benchmark for several downstream tasks requiring pre-trained sequence embeddings. 
% for fun
% \begin{equation}
%  l_{r} = \frac{\texttt{dist}(h(r), h(\mathcal{F}_j))}{\tau} - \log\left(1 + \sum^{B-1}_{j=1} e^{-\texttt{dist}(h(r), h(\mathcal{F}_j))/\tau}\right) 
% \end{equation}

\subsection{2.4 RDE: Encoder implementation}

RDE uses a Transformer-encoder \citep{devlin_bert_2019, vaswani_attention_2017}, comprising $12$ heads and $6$-layers of encoder blocks. The size of the vocabulary is $10,000$. Batch size $B$ is set to $16$ with gradient accumulation across $16$ steps.  Generated numerical representations for each read (and fragment) are projected into a high-dimensional vector space, $h(r) \in \mathbb{R}^{1020}$. The learning rate is annealed using one-cycle cosine annealing \citep{smith_super-convergence_2019}, dropout is set to $0.1$, and $\tau=0.05$. During training, the reference fragment $|\mathcal{F}_i| \sim \mathcal{U}([800, 2000])$ and
read $|r_i| \sim \mathcal{U}([150, 500])$ have variable lengths sampled from a uniform distribution. Here, $|x|$ denotes the length of $x$. To improve model performance on noisy reads, $1-5\%$ of bases are replaced with another random base in $40\%$ of the reads in a batch. Shorter sequences were padded to equal the length of the longest sequence in a batch. The similarity measure used is \textit{Cosine Similarity}.

\subsection{2.5 ESA: Search and retrieval} \label{sandr}

% \begin{figure}
%     \centering
%     \includegraphics[width=0.9\columnwidth]{images.py/down4.pdf}
%     \caption{\textbf{System Overview [B] - Inference on a New Read:} A read, as per \eqref{eq:read} and generated by ART~\citep{huang_art_2012}, is encoded by $h$. This is then compared to reference fragment representations in the vector database. The nearest-$K$ fragments in the embedding space are retrieved for each read, and the optimal alignment is determined using \eqref{eq:final}.
%     }
%     \label{fig:down}
%     \vspace{-1em}
% \end{figure}

% \begin{figure}
%     \centering
%     \includegraphics[width=\columnwidth,height=0.9\textheight,keepaspectratio]{images.py/down4.pdf}
%     \caption{\textbf{System Overview [B] - Inference on a New Read:} A read, as per \eqref{eq:read} and generated by ART~\citep{huang_art_2012}, is encoded by $h$. This is then compared to reference fragment representations in the vector database. The nearest-$K$ fragments in the embedding space are retrieved for each read, and the optimal alignment is determined using \eqref{eq:final}.
%     }
%     \label{fig:down}
%     \vspace{-1em}
% \end{figure}

\begin{figure*}[t]
    \centering
    \includegraphics[width=0.8\textwidth]{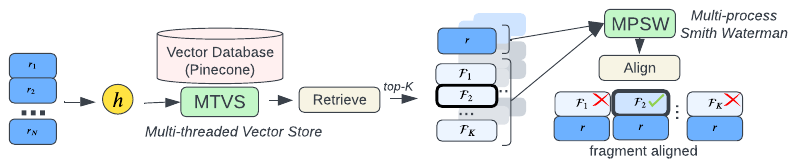}
    \caption{\textbf{System Overview [B] - Inference on a New Read:} A read, as per \eqref{eq:read} and generated by ART~\citep{huang_art_2012}, is encoded by $h$. This is then compared to reference fragment representations in the vector database. The nearest-$K$ fragments in the embedding space are retrieved for each read, and the optimal alignment is determined using \eqref{eq:final}.}
    \label{fig:down}
\end{figure*}

An outline of the search and retrieval process is presented in Fig.~\ref{fig:down}. Every read is encoded using the trained model and matched to reference fragments in the vector database. The top-$K$ retrieved fragments per read are then aligned using a SW alignment library to find the optimal alignment. The following sections describe the indexing and retrieval part in more detail.

% \subsubsection{Indexing} \label{sec:indexing}

\phantomsection{\textbf{Indexing:} For a given reference genome $\mathcal{R}$, we construct a minimal set of reference fragments $\mathcal{F} := \{ \mathcal{F}_1, \mathcal{F}_2, \dots \}$ to span $\mathcal{R}$. Note that the fragments overlap at least a read length; i.e. $|\mathcal{F}_i \bigcap \mathcal{F}_{i+1}| \ge Q$ to guarantee that every read is fully contained within some fragment in the set. In our experiments with external read generators~\citep{huang_art_2012}, $Q_{max} = 250, |\mathcal{F}_i| = 1250$. Each reference fragment is encoded using the trained \textit{RDE} model, and the resulting sequence embeddings ($\in \mathbb{R}^{1020}$) -- $3$M vectors for a reference of $3$B nucleotides -- are inserted into a Pinecone\footnote{\url{https://www.pinecone.io}} database. Once populated with all the fragments, we are ready to perform the alignment.}\label{sec:indexing}

\phantomsection{\textbf{Retrieval:} Given a read $r$, we project its corresponding \textit{RDE} representation into the vector store and retrieve the approximate nearest-$K$ set of reference fragment vectors and the corresponding fragment metadata $\{\mathcal{F}_1, \mathcal{F}_2, \dots, \mathcal{F}_K\}$.} \label{sec:retrieval} 

\phantomsection{\textbf{Diversity priors:} While the top-$K$ retrieved fragments can be drawn from across the entire vector store (genome), contemporary recommendation systems that use the top-$K$ retrieval setup \textit{rank and re-rank} top search results (\textit{Slate Optimization} -- see \cite{grasshopper}) to ensure rich and diverse recommendations. Similarly, we apply a uniform prior wherein every retrieval step selects the top-$K$ \textit{per} chromosome.} \label{sec:diversitypriors}

\textbf{Fine-Alignment:} A standard SW score library~\citep{biopython} is used to solve~\eqref{eq:main_opt}, which can be executed concurrently across the $K$-reference fragments. Let the optimal fragment be $\mathcal{F}^*$. The metadata for each vector includes (a) the raw $\mathcal{F}^*$ sequence; (b) the start position of $\mathcal{F}^*$ within the reference $\mathcal{R}$, $q_{\mathcal{F}^* | \mathcal{R}}$. Upon retrieval of a fragment and fine-alignment to find the fragment-level start index, $q_{| \mathcal{F}^*}$, the global reference start position is obtained as:
\begin{equation}
q^* = q_{| \mathcal{F}^*} + q_{\mathcal{F}^* | \mathcal{R}}.
\label{eq:final}
\end{equation}

\section{3. Transformer-DNA baselines} \label{sec:alignpure}
% We detail the testbench setup for evaluating the existing Transformer-DNA baselines whose recall curves are illustrated in Fig.~\ref{fig:puretests}. Three recent architectures that encode sequences at the nucleotide level are selected: [NT] \href{https://huggingface.co/InstaDeepAI/nucleotide-transformer-500m-human-ref}{\textit{NucleotideTransformer}} ($\in \mathbb{R}^{1280}$)~\citep{nucleotide}; [DB2] \href{https://huggingface.co/zhihan1996/DNABERT-2-117M}{\textit{DNABERT-2}} ($\in \mathbb{R}^{768}$)~\citep{ji_dnabert_2021}; [HD] \href{https://huggingface.co/LongSafari/hyenadna-small-32k-seqlen}{\textit{HyenaDNA}}($\in \mathbb{R}^{256}$)~\citep{nguyen_hyenadna_2023}. For each of the NT, DB2, HD architecture, a mean- and-max pooling ($2$) of the token representations yields the sequence representation ($2 \times 3 = 6$ baselines total). An independent vector store is used for each of the six baselines; i.e. reference fragments from the entire genome of $3$gb length are encoded using each of the $6$ encoders (see Sec.~\ref{sandr}); $40$K pure reads -- reads without mutations and variations -- of length $Q \sim \mathcal{U}[25,1000]$ ($x$) are sampled from across the reference and the average recall ($y$) for top-$5$, top-$25$, and top-$50$ retrieved fragments -- without diversity priors -- is reported with respect to read length (see Fig.~\ref{fig:puretests}). While the baselines perform poorly across the board, mean-pooling performs better than max-pooling (except for HD). And DB2 (mean-pooled) and HD (max-pooled) perform better than the rest. We will use these two baselines in Table~\ref{tab:mainres} to compare with DNA-ESA.

This section outlines the setup for evaluating Transformer-DNA baselines, with their recall performance depicted in Fig.~\ref{fig:puretests}. We selected three architectures modeling nucleotide sequences: [NT] \textit{NucleotideTransformer} ($\in \mathbb{R}^{1280}$)~\citep{nucleotide}, [DB2] \textit{DNABERT-2} ($\in \mathbb{R}^{768}$)~\citep{ji_dnabert_2021}, and [HD] \textit{HyenaDNA} ($\in \mathbb{R}^{256}$)~\citep{nguyen_hyenadna_2023}. Each model employs mean- and max-pooling of token representations for sequence encoding ($2 \times 3 = 6$ baselines total). Independent vector stores for each baseline encode fragments from the entire $3$gb genome. We sampled $40$K pure reads (reads without noise in comparison to the reference) of varying lengths ($Q \sim
\mathcal{U}([25,1000])$) and assessed the average recall for top-$5$, top-$25$, and top-$50$ fragments, as shown in
Fig.~\ref{fig:puretests}. Overall, while baseline performance is modest, mean-pooling generally outperforms max-pooling, with DB2 (mean-pooled) and HD (max-pooled) as the most effective. These two baselines will be contrasted with RDE on ESA in Table~\ref{tab:mainres}.

RDE convergence plots are presented in the SM Sec. A. Model checkpoints are available in OSF~\cite{osf}. In Fig.~\ref{fig:eyecandy}, representations of short $1,000-$length sequences sampled from sequential (in-order) and gene-specific locations in the reference are visualized in a reduced 2D-UMAP~\citep{umap}. The representation space demonstrates desired properties suitable for successfully performing alignment: (a) Sequences sampled in order form a trajectory in the representation space: The loss function described in \eqref{eq:loss} encourages a pair of sequences \textit{close} to one another to have a short distance between them in the representation space, and pairs further apart to have a larger distance. (b) Representations of sequences drawn from specific gene locations -- despite not being close to one another -- show gene-centric clustering: The RDE representation space partially acquires function-level separation as a byproduct of imposing \textit{local} alignment constraints. The codebase is \href{https://anonymous.4open.science/r/dna-esa-BED6/readme.md}{\textit{linked}}.

\section{4. Sequence alignment of ART-simulated reads} \label{sec:seqa}

The results from Sec.~\ref{sec:alignpure} demonstrate that even for pure reads, baseline models do not generate adequate representations to perform sequence alignment. In this section, RDE and the two best baselines---\textit{DB2, mean} and \textit{HD, max}---are evaluated on ESA using reads generated from an external read simulator (ART) -- see \cite{huang_art_2012}. ART has served as a reliable benchmark for evaluating other contemporary alignment tools and provides controls to model mutations and variations common in reads generated by Illumina machines.

\noindent \textbf{Simulator configurations:} The different simulation configuration options and settings are listed: (A) \textit{Phred quality score} $Q_{PH}$ in one of three $\{[10,30], [30,60], [60,90]\}$ ranges: the likelihood of errors in base-calls of a generated read; (B) \textit{Insertion rate} $I \in \{0, 10^{-2}\}$: the likelihood of adding a base to a random location in a read; (C) \textit{Deletion rate} $D \in \{0, 10^{-2}\}$: the likelihood of deleting a base in the read; (Others): \textit{Simulator system}: MSv3 [MiSeq]; \textit{Read length}: $250$.

\noindent \textbf{Recall configurations:} Once the top-K fragments have been retrieved, the first step is to solve \eqref{eq:main_opt}, and for this, we need to compute the SW score. For all presented results, the settings are: \texttt{match\_score} = -2, \texttt{mismatch\_penalty} = +1, \texttt{open\_gap\_penalty} = +0.5, \texttt{continue\_gap\_penalty} = +0.1. After alignment, we get $q^*$---see \eqref{eq:final}---as the estimated location of a read in the genome. Let $\hat{q}^*$ be its true location. If $q^* = \hat{q}^*$, it is a perfect match and the recall is successful. In cases where there is a mutation in the first or last position in a read, the fine-alignment will return $q^*_{\mathcal{F}^*}$ offset by at most $2$ locations, resulting in $\hat{q}^* = q^* \pm 2$. Hence, the condition for an exact location match: $|q^* - \hat{q}^*| \leq 2$.

\noindent \textbf{Distance bound}, $d_{SW} \in \{1\%, 2\%, 5\%\}$: It is well known that short fragments frequently repeat in the genome and $q^*$ can correspond to the position of the read in a different location than from where it was sampled~\cite{mappable, reinventwillis}. In this case, $q^* \neq \hat{q}^*$, but the SW score is the minimum possible. Moreover, when reads have mutations, the reference sequence corresponding to the read is no longer a perfect match. We define a bound $d_{SW}$ for classifying whether a (read, retrieved fragment) pair is a successful alignment based on the SW score between the pair ($v^*$ -- see Eq.~\ref{eq:main_opt}) and the optimal SW score as a particular read length $Q$:

$$
d_{SW} = \frac{v^* - mQ}{(n-m)Q},
$$
where $m(=-2)$ is the match score and $n(=+1)$ is the mismatch penalty, hyperparameters in the computation of the SW score. We consider an alignment (with $Q=250$) to be successful if the resulting SW score between the read and the top returned fragment is within $d_{SW}$ of the optimal for that read length. A $d_{SW} = 2\%$ (\textit{default}) is equivalent to a mismatch of around $4$ bases in a read of length $250$.

\begin{table*}[ht]
\centering
\renewcommand{\arraystretch}{0.6}
\begin{tabular}{@{}c|lllll|l@{}}
\toprule
\textbf{Model} & \multicolumn{1}{l|}{\textbf{Settings}} & \multicolumn{4}{c|}{\textbf{\begin{tabular}[c]{@{}c@{}}ART\\ (MiSeqv3)\end{tabular}}} &

 \multicolumn{1}{c}{\begin{tabular}[c]{@{}c@{}}\textbf{PacBio CCS}\\($\mu(v^*), \sigma(v^*)$)\end{tabular}}

\\ \midrule
\textbf{} & \multicolumn{1}{l|}{\begin{tabular}[c]{@{}l@{}}$I,D = 0.01, Q=250$,\\ $d_{SW}=2\%$\end{tabular}} & \multicolumn{1}{l|}{Search} & \multicolumn{1}{l|}{{[}10,30{]}} & \multicolumn{1}{l|}{{[}30,60{]}} & {[}60,90{]} & \multicolumn{1}{c}{\begin{tabular}[c]{@{}c@{}}Chr. 2, pbmm2\\ $d_{SW}=5\%$\end{tabular}} \\ \midrule
\multirow{4}{*}{\begin{tabular}[c]{@{}c@{}}Transformer \\ Architectures\end{tabular}} & \multicolumn{1}{l|}{HD (max)} & \multicolumn{1}{l|}{\multirow{3}{*}{K=50}} & \multicolumn{1}{l|}{13.20 $\pm$ 1.81} & \multicolumn{1}{l|}{17.01 $\pm$ 2.01} & 18.03 $\pm$ 2.05 & \multicolumn{1}{c}{-} \\ \cmidrule(lr){2-2} \cmidrule(l){4-7} 
 & \multicolumn{1}{l|}{DB2 (mean)} & \multicolumn{1}{l|}{} & \multicolumn{1}{l|}{30.21 $\pm$ 2.44} & \multicolumn{1}{l|}{39.19 $\pm$ 2.58} & 39.20 $\pm$ 2.58 & \multicolumn{1}{c}{-} \\ \cmidrule(lr){2-2} \cmidrule(l){4-7} 
 & \multicolumn{1}{l|}{\multirow{2}{*}{\textbf{\begin{tabular}[c]{@{}l@{}}RDE\\ (ours)\end{tabular}}}} & \multicolumn{1}{l|}{} & \multicolumn{1}{l|}{98.40 $\pm$ 0.71} & \multicolumn{1}{l|}{98.64 $\pm$ 0.67} & 98.60 $\pm$ 0.67 & \multicolumn{1}{c}{97.5 $\pm$ 0.82}\\ \cmidrule(l){3-7} 
 & \multicolumn{1}{l|}{} & \multicolumn{1}{l|}{K=75} & \multicolumn{1}{l|}{\textbf{98.80 $\pm$ 0.62}} & \multicolumn{1}{l|}{\textbf{99.28 $\pm$ 0.52}} & \textbf{98.88 $\pm$ 0.60} & 
 
 \multicolumn{1}{c}{\begin{tabular}[c]{@{}c@{}}\textbf{97.5 $\pm$ 0.82}\\ (-498.0, 3.78)\end{tabular}}

 \\ \midrule
\begin{tabular}[c]{@{}c@{}}Bowtie-2 \\ (Classical)\end{tabular} & \multicolumn{2}{c|}{-} & \multicolumn{3}{c|}{99.80 $\pm$ 0.19} & 

 \multicolumn{1}{c}{\begin{tabular}[c]{@{}c@{}}99.50$\pm$0.43\\ (-497.9, 3.78)\end{tabular}}

\\ \midrule
\textit{diff.} & \multicolumn{2}{c|}{-} & \multicolumn{3}{c|}{\textless{}1\%} & \multicolumn{1}{c}{\textless{}2\%} \\ \bottomrule
\end{tabular}
\vspace{1em}
\caption{
\textbf{Performance of RDE with respect to baselines:} The performance of RDE on ESA is compared to the DNA-BERT 2 and Hyena-DNA Transformer-based baseline models (with no additional finetuning) -- the top performing baselines from Fig.~\ref{fig:puretests}. In addition, RDE is also compared with Bowtie-2~\cite{bowtie}, a classical aligner. Comparisons are conducted across varying qualities (\textit{Phred score}) of reads generated by ART~\cite{huang_art_2012}.  Additionally, an external PacBio CCS dataset of ($Q=250$ length-) reads from Chr. 2 of the Ashkenazim Trio - Son sample (as determined by pbmm2~\cite{minimap}) are aligned using both Bowtie-2 and RDE. All baseline models utilize a dedicated vector store. For details on $I, D, Q_{PH}, K, d_{SW}$, refer to \textit{Recall/Simulator Configurations}. Across models and ART-generated datasets, the performance of RDE supersedes other Transformer-based models and is comparable to Bowtie-2 to within $1\%$. With respect to reads from PacBio CCS, RDE performs with $2\%$ of Bowtie-2 after controlling for the intrinsic quality of retrieved fragments according to the SW score. Mean ($\mu$) and standard deviation ($\sigma$) across the top SW scores($v^*$) of reads successfully aligned in the case of RDE and Bowtie-2 models are reported.
}
\label{tab:mainres}
\end{table*}

\textcolor{black}{\subsection{4.1 Performance}\label{app:noisy} 
Table~\ref{tab:mainres} reports the performance of RDE on ESA across several read generation configurations -- we choose from one of three ranges of Phred score $Q_{PH}$, $\{[10,30],[30,60],[60,90]\}$ -- $I,D = 0.01$, $d_{SW} = 2\%$ -- see Sec.~\ref{sec:seqa}), in addition to a direct comparison to \texttt{DB2, mean} and \texttt{HD, max} baselines (\textit{without any further finetuning}), the best-performing baselines on pure reads identified in Sec.~\ref{sec:alignpure}. In addition, we also compare RDE to Bowtie-2~\cite{bowtie}, a conventional algorithmic aligner.}

\textcolor{black}{Additionally, models are also evaluated on an external set of reads from PacBio CCS generated on the Ashkenazim Trio (Son) (retrieved from the Genome in a Bottle resource (GIAB)~\cite{giab}). For (ii), the raw reads are $10$ kilobases long and for evaluation we consider random subset of $5000$ reads associated to Chr. 2. Reads input into the aligner are cropped randomly to a length of $250$ to satisfy the computational requirements of RDE ($|\mathcal{F}_i \bigcap \mathcal{F}_{i+1}| = 250$).}

\textcolor{black}{We observe that: (A) RDE shows strong recall performance of $> 99\%$ across a variety of read generation and recall configurations; (B) On cleaner reads, RDE and Bowtie-2 are within $<1\%$ of each other in recall while controlling for the quality of the retrieved alignments (as determined by the SW score). The results indicate that this constitutes a new state-of-the-art model for Transformer-based sequence alignment.}

% \textcolor{black}{
% \noindent \textbf{On high-noise reads:} 
% In Table~\ref{tab:highnoise}, we report the performance of DNA-ESA on two noisy datasets: (i) ART-generated (MiSeqv3-based) reads of length $Q=250$ with Phred quality score $Q_{PH} \in [10,20], I,D = 1\%$ and (ii) reads from PacBio CCN as generated on the Ashkenazim Trio (Son) (retrieved from the Genome in a Bottle resource (GIAB)~\cite{giab}). For (ii), the raw reads are 10 kilobases long and for evaluation we consider random subset of $1000$ reads derived to Chr. 2. Reads specific to a chromosome are filtered using the pbmm2 (Minimap) package~\cite{minimap}. Reads input into the aligner are cropped to length $250, 350$ and $500$. Note that in these experiments, we take care to increase the fragment-to-fragment overlap $|\mathcal{F}_i \bigcap \mathcal{F}_{i+1}|$ from $250$ to $500$. DNA-ESA performance is reported with respect to the Bowtie-2 performance. DNA-ESA performs comparably to Bowtie-2 ($<1\%$ difference in the recall controlling for the quality of the successfully aligned reads (in DNA-ESA and Bowtie-2) by using the corresponding Smith-Waterman distances).}

%\input{tables.py/noisy}

\textcolor{black}{
\noindent \textbf{Sweeping Top-K and the SW distance bound} \label{app:topk}
In Table~\ref{tab:suppres}, we report the performance of RDE across distance bounds $d_{SW} \in \{1\%, 2\%, 5\%\}$ and a number of recalled fragment settings $K \in \{25,50,75\}$. Distance bound $d_{SW} < 5\%$ is equivalent to an acceptable mismatch of at most $\sim 8$ bases between the read (length $Q=250$) and recalled reference fragments. Recall rates are comparable with different $d_{SW}$ values suggesting that when a read is successfully aligned, it is usually aligned to an objectively best match. Decreasing the top-$K$ per chromosome from $50 \rightarrow 25$ does not substantially worsen performance ($<1\%$), indicating that the optimal retrievals are usually the closest in the embedding space. \textcolor{black}{Several additional experiments are presented in the SI.}}

\begin{table*}[ht]
\centering
\renewcommand{\arraystretch}{0.7}
\begin{tabular}{ll|ccc|ccc}
\toprule
 &  & \multicolumn{3}{c|}{$Q_{PH} \in [30, 60]$} & \multicolumn{3}{c}{$Q_{PH} \in [60, 90]$} \\
$I$ & $D$ & $d_{SW} < 1\%$  & $d_{SW} < 2\%$ & $d_{SW} < 5\%$ & $d_{SW} < 1\%$  & $d_{SW} < 2\%$ & $d_{SW} < 5\%$  \\
\midrule
\midrule
 \multicolumn{8}{c}{Top-25 / chromosome} \\
\midrule
\multirow[t]{2}{*}{0.0} & 0.0 & 97±0.94 & 98.12±0.76 & 98.56±0.69 & 97.24±0.91 & 97.96±0.0.80 & 98.6±0.69 \\
 & 0.01 & 96.88±0.97 & 98±0.78 & 98.36±0.73 & 97.24±0.91 & 97.96±0.80 & 98.76±0.64 \\
\multirow[t]{2}{*}{0.01} & 0.0 & 97.08±0.93 & 98.04±0.78 & 98.56±0.69 & 97±0.94 & 98.24±0.75 & 98.92±0.59 \\
 & 0.01 & 97.16±0.80 & 97.96±0.56 & 99.2±0.52 & 97.6±0.85 & 98.12±0.76 & 98.76±0.62 \\
  \midrule 
 \multicolumn{8}{c}{Top-50 / chromosome} \\
 \midrule
\multirow[t]{2}{*}{0.0} & 0.0 & 97.52±0.86 & 99.04±0.57 & 99.08±0.57 & 98.2±0.75 & 98.88±0.59 & 99.16±0.52 \\
 & 0.01 & 97.48±0.87 & 99.16±0.52 & 99.08±0.55 & 97.8±0.82 & 98.76±0.62 & 99.04±0.57 \\
\multirow[t]{2}{*}{0.01} & 0.0 & 98.24±0.75 & 98.64±0.67 & 99.12±0.55 & 98.28±0.75 & 98.92±0.59 & 99.28±0.49 \\
 & 0.01 & 97.72±0.83 & 98.64±0.67 & 99.36±0.46 & 97.84±0.92 & 98.6±0.66 & 99.0±0.57 \\
\midrule
\multicolumn{8}{c}{Top-75/ chromosome} \\
 \midrule
\multirow[t]{2}{*}{0.0} & 0.0 & 98.04±0.78 & 99.12±0.55 & 99.0±0.57 & 98.4±0.71 & 99.12±0.55 & 98.4±0.71 \\
 & 0.01 & 98.4±0.71 & 99.04±0.56 & 99.4±0.46 & 98.44±0.71 & 98.84±0.62 & 99.24±0.52 \\
\multirow[t]{2}{*}{0.01} & 0.0 & 98.64±0.67 & 99.0±0.57 & 99.2±0.52 & 98.0±0.78 & 98.72±0.64 & 99.48±0.43 \\
 & 0.01 & 98.68±0.64 & 99.28±0.49 & 99.4±0.46 & 98.48±0.69 & 98.88±0.59 & 99.4±0.46 \\
\bottomrule
\end{tabular}
\vspace{1em}

\caption{\textbf{Sequence alignment recall of RDE sweeping top-K and $d_{SW}$:} The various parameters are described in Sec.~\ref{sec:seqa}. RDE presents a recall of $>99\%$ across several read configurations rivaling contemporary algorithmic models such as Bowtie. As expected, performance improves with larger search radius (top-$K$), higher quality reads ($Q_{PH}$) and large distance bound $d_{SW}$.}
\label{tab:suppres}
\end{table*}

\section{5. Limitations and Future Work}
While our first-generation reference-free DNA embedding (RDE) models provide alignment performance that almost matches the performance of traditional aligners (refined over decades), there is room for considerable improvement. First, our RDE models have been trained using samples that span only around 2\% of the human genome as shown in the Sec-A of the Supporting Material(SM). One needs to explore if the embedding properties of our models can be further improved by diversifying and augmenting training data and optimizing over the model parameters. The validation of RDE models can be performed only through various genomic tasks. In terms of the task of alignment, we plan to explore the following: (i) The current off-the-shelf implementation of our ESA algorithm has a speed of around 10K reads per minute. The existing aligners such as Bowtie are faster, and  can achieve close to 1M reads per minute; additional speedups can be achieved  by trading off alignment accuracy \cite{vargas}. \color{black}   As detailed in SM Sec. B, we are considering various optimization strategies, including model compilation, to speed up inference and enhanced parallelization in vector store searches and fragment-read alignment. We believe ESA can be made much faster; (ii) Optimization of the training of the RDE models to improve performance for shorter reads. As shown in see Table 2, Page 3 of SM the existing model performs well on longer reads; we want to develop RDE models that specialize to aligning shorter reads; and (iii) quantifying how embeddings change as subsequences are edited; this will help in better alignment accuracy. It will also help in correlating the geometry of the embedding space to the structure of DNA sequences.

\section{6. Concluding Remarks}

We have introduced a novel Reference-Free DNA Embedding (RDE) framework: it creates an embedding of any given DNA subsequence irrespective of the reference genome from which it is sampled. Such models capture similarity among genome subsequences in a \textit{purely data-driven manner} (without explicitly computing edit-distance measures) and embeds DNA sequences of differing lengths into the same embedding space to facilitate search and assembly. Once such RDE models are constructed their expressive power and limitations can be assessed by  performing multiple challenging genomic tasks in a flexible manner across different species.  

As a first validation and application of RDE, we developed an Embed-Align-Search (ESA) framework where alignment of reads to a reference genome is achieved by a near-neighbor search in the embedding space: only the fragments of reference genome with embeddings nearest to the embedding of the given read need to be considered. Thus, a global search in a giga-bases long reference genome sequence is reduced to a local search in a vector space. Detailed experiments showed that the performance of even this first-generation RDE methodology is comparable to that of traditional aligners, which have been refined over decades. Future application of RDE to the alignment task might lie in aligning reads to pan-genomes~\cite{liao}, where the diversity in the genomic content -- modeled as allelic paths in a reference genome graph -- can be exhaustively mapped onto the described embedding space (as in Fig. ~\ref{fig:eyecandy}). Conveniently, the proposed nearest-neighbor search would align a given read to fragments; the only difference being that multiple fragments can belong to the same reference location but different reference paths. The ESA methodology would not need to be modified beyond this. \color{black}

% Thus, the ESA methodology does not have to be significantly altered, and alignment of reads to pangenomes constitutes an important future task. 

% Indexing such pangenomes is a difficult task in traditional aligners; however,

An important question we want to answer going forward is: How reference-free is our RDE model? To benchmark the generalizability of our RDE model we have performed several  cross-species alignment tasks as summarized in Tables  3 and 4  in the SM. In particular, we have used an RDE trained using human genome to align reads to reference genomes belonging to a wide range of species. The results are encouraging and the design of universal RDEs---that would generate effective embeddings of subsequences sampled from the genome of every species---is another direction of future research. 

%Current DNA sequence alignment methods, honed over decades, incorporate DNA-specific enhancements in both indexing and retrieval, relying on algorithmic approaches. However, we introduce an alternative data-driven paradigm, employing a Transformer-based DNA-ESA encoder.  Our method performs sequence alignment through self-supervised learning, utilizing contrastive loss. While similar methods have previously found application in identifying approximate semantic similarity in written language, our approach surprisingly excels at identifying exact overlaps among DNA sequences. Empirical results underscore the model's ability to embody the inherent structure of any DNA sequence, irrespective of location or species origin. A distinguishing feature of DNA-ESA is its ability to map both shorter reads and longer reference fragments into the same space by employing cosine similarity as a symmetric distance metric. This complements existing paradigms that have traditionally used a hierarchical and explicit comparison of base distributions.  

%DNA-ESA representations would find utility in other related tasks. If, for example, one had to align reads to the Pan Genome~\cite{liao} instead of a single reference, one can populate a single vector store with fragments from all $47$ genome assemblies. Then given a read, the nearest fragment to the read can belong to any reference variant. Note that in this case, we avoid having to construct a separate index for each of the references as is usually done in conventional methods.

One can also harness the RDE space  to address the complementary task of \textit{de novo Genome Assembly}, where given a set of reads one assembles a genome by stitching together the reads, without using any reference. Genome assembly is considered to be a much more challenging task than alignment. In SM Sec. F, we provide initial proofpoints of a genome assembly framework, under reasonable  simplifying assumptions about read lengths and overlaps among the reads. We show how the RDE model---trained on exemplars from human genome---can effectively embed simulated reads from a very different species, Thermus Aquaticus. We find that  the read embeddings form a low-dimensional manifold, which can be efficiently searched locally to obtain high-fidelity genome assembly from the simulated reads. Our future work on genome assembly will focus on improving RDE models to obtain better embeddings, and also on developing local-search based algorithms that can accurately map low-dimensional manifolds (comprised by the embeddings of reads) into linear DNA sequences.

\bibliographystyle{plain}
\bibliography{reference}

\begin{appendices}
    
\section{Appendix}

\section{A. Training Convergence and Data Usage} \label{app:convergence}
\begin{figure}[h]
    \centering
    \begin{subfigure}{0.48\columnwidth}
        \includegraphics[width=\textwidth]{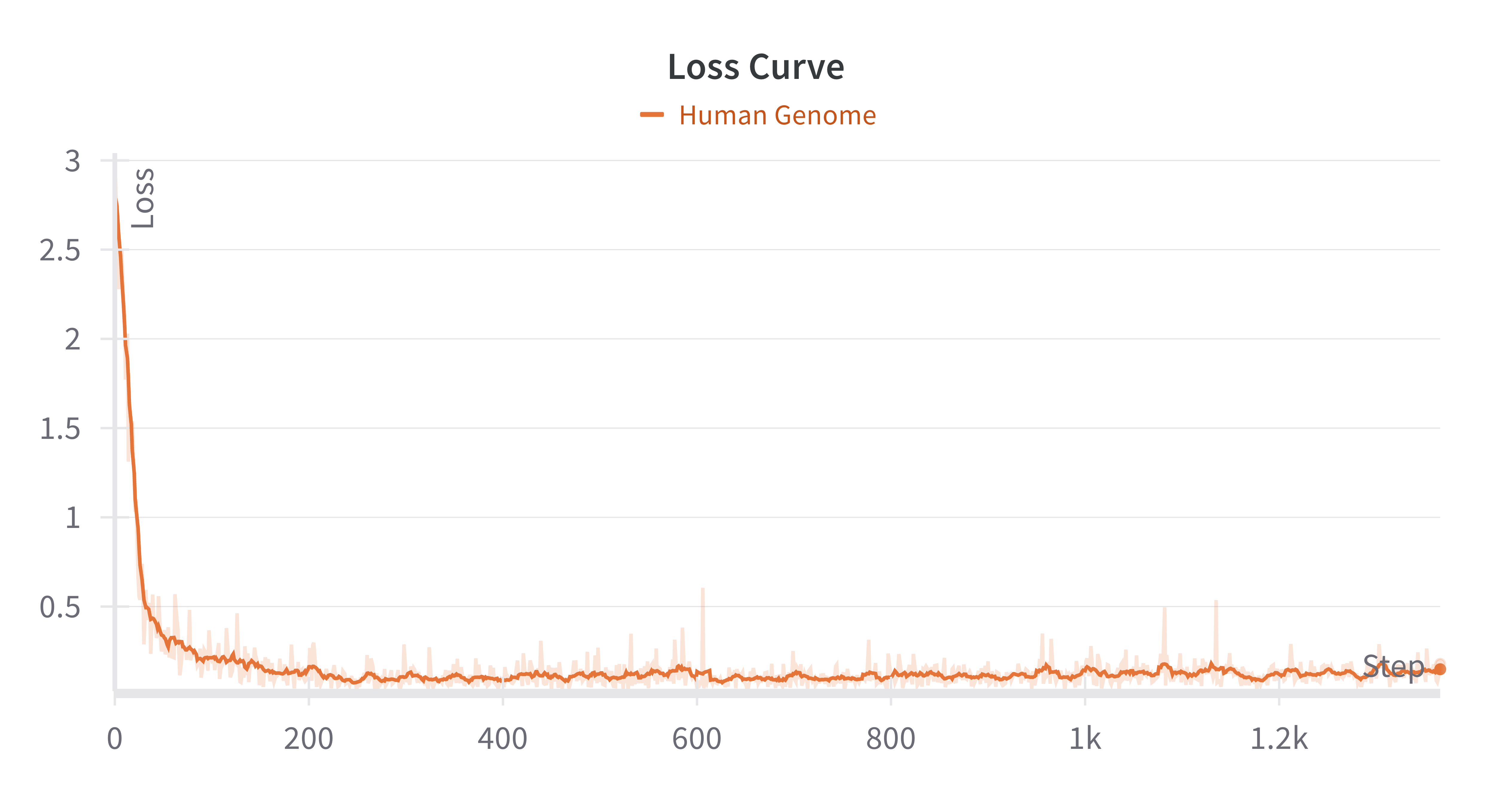}
        \caption{Trained on the Human Genome}
        \label{fig:loss1}
    \end{subfigure}
    \hfill
    \begin{subfigure}{0.48\columnwidth}
        \includegraphics[width=\textwidth]{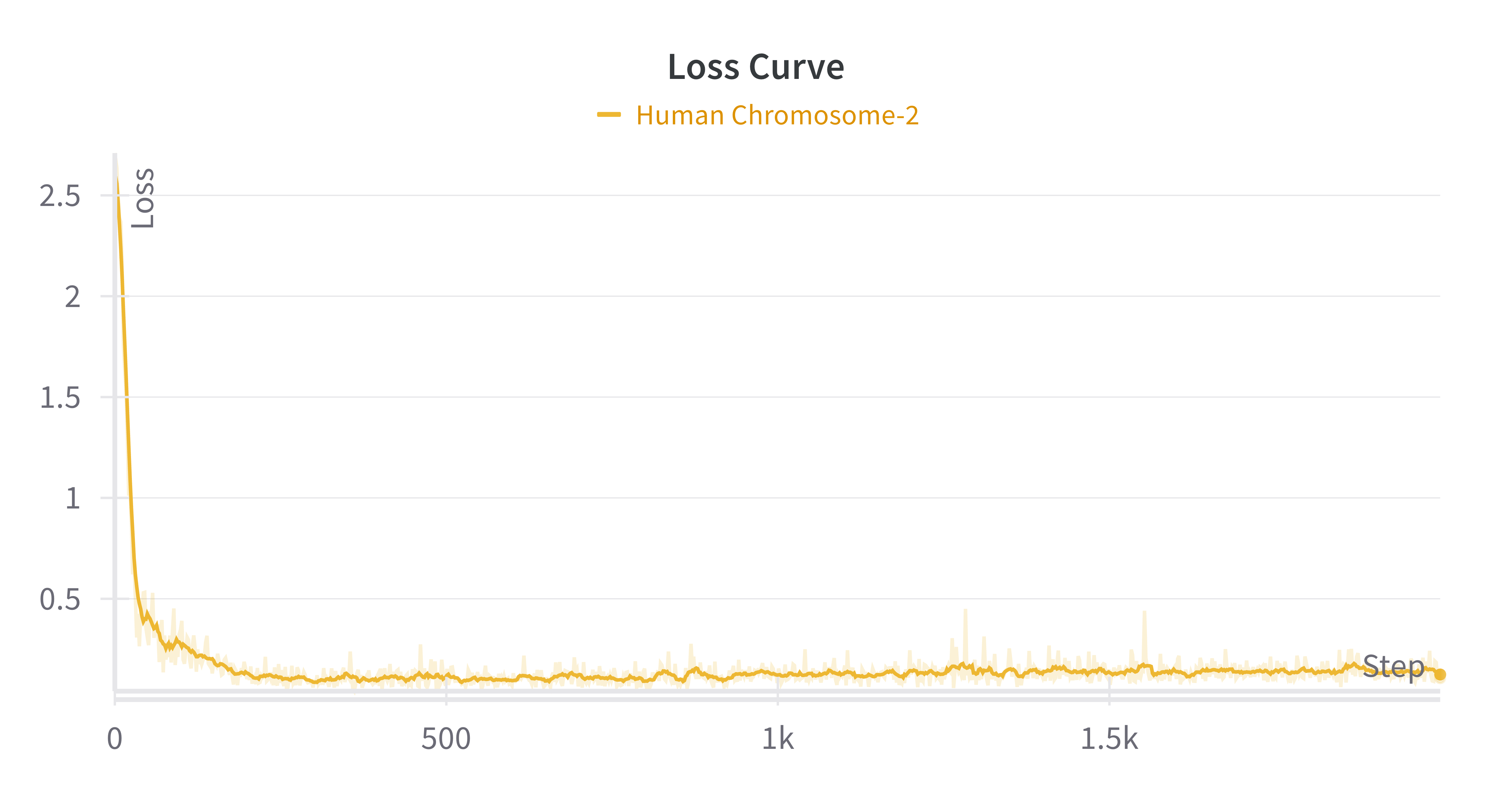}
        \caption{Trained on the Human Chromosome 2}
        \label{fig:loss2}
    \end{subfigure}
    \caption{RDE convergence plots. Both plots show the loss pr. step. For clarity, we smooth the loss using a moving average}
    \label{fig:loss}
\end{figure}

Fig.~\ref{fig:loss} plot the convergence of the RDE encoder model discussed in the main text. We see convergence after $\sim$2k steps.

% \begin{itemize}
%     \item With 2000 steps and a batch size of 16, we have a total of $$2000 \times 16 = 32,000$$ training fragments.
%     \item The maximum size of a fragment is 2000 base pairs.
%     \item Therefore, the total number of base pairs seen during training is approximately $$32,000 \times 2000 = 64000000$$ (64 million) base pairs.
%     \item The human genome consists of around 3 billion base pairs.
%     \item Consequently, our model sees only about 2\% (64 million / 3 billion) of the entire human genome during training.
% \end{itemize}
 With 2000 steps and a batch size of 16, we have a total of $2000 \times 16 = 32,000$ training fragments. The maximum size of a fragment is 2000 base pairs. Therefore, the total number of base pairs seen during training is approximately $32,000 \times 2000 = 64000000$ (64 million) base pairs.The human genome consists of around 3 billion base pairs. Consequently, our model sees only about 2\% (64 million / 3 billion) of the entire human genome during training.

This limited exposure to the full genome highlights the RDE model's ability to generalize from a relatively small subset of the available data. Despite seeing only a fraction of the human genome, RDE demonstrates strong performance in sequence alignment tasks across various chromosomes and even across species.

% \begin{figure}[h]
%     \centering
%     \includegraphics[width=0.5\textwidth]{images.py/loss.png}
%     \caption{\textbf{Convergence plot:} Plot demonstrates stable convergence after $\sim$20 thousand iterations of the DNA-ESA encoder trained on the whole human genome.}
%     \label{fig:loss}
% \end{figure}

% \begin{figure}[h]
%     \centering
%     \includegraphics[width=0.5\textwidth]{images.py/rich-frog-63-loss.png}
%     \caption{\textbf{Convergence plot:} Plot demonstrates stable convergence after $\sim$20 thousand iterations of the DNA-ESA encoder trained on the Human Chromosome 2.}
%     \label{fig:loss}
% \end{figure}

\section{B. Complexity of computing alignment} \label{app:complexity}
\subsection{Cost of constructing a new representation}

Computing the embedding $\mathcal{E}$ of a sequence of length $F$ using RDE encoding -- a typical Transformer-based attention architecture -- has the following computation complexity:
\begin{equation}
    \mathcal{O}(LH*(F^2*d + d^2*F)) \Rightarrow \mathcal{O}(F^2*d + d^2*F) 
\end{equation}

Where $d$ is the embedding dimension of the model, $L$ is the number of layers in the Transformer and $H$ is the number of heads per layer. As $d$ is a controllable parameter for the model, we can further simplify:
\begin{equation}
\mathcal{O}(F^2*d + d^2*F) \Rightarrow  \mathcal{O}(F^2)
\end{equation}

The $F^2$ complexity follows the basic implementation of attention in transformers, but recent efforts~\citep{beltagy_longformer_2020, kitaev_reformer_2020} 
have developed shortcuts to reduce the cost. These have already been applied to DNA sequence modeling ~\citep{nguyen_hyenadna_2023}.

\subsection{Vector Store Upstream}

Vector store $\mathcal{D}$ is populated once (in bulk) with encoded fragment-length sequences drawn from the entire genome; \textit{constant time complexity} $C$ to upload $<10M$ vectors.

\subsection{Retrieval Cost}

Given a new embedding, $\epsilon(G, K)$ is the cost of retrieving top-$K$ nearest neighbors across the fragment embeddings, where $G$ is the length of the reference genome. In modern vector databases, where several hashing techniques such as approximate K-nearest neighbors are used, $\epsilon$ scales logarithmically with $G$. This is indeed the key benefit of using such databases.

\subsection{Fine-grained Alignment}

Existing libraries/algorithms (e.g. the Smith–Waterman algorithm) can identify the alignment between a fragment sequence (of length $F$) and read (of length $Q$) in $\mathcal{O}(FQ)$.

\subsection{Total Complexity}

Total complexity involves (a) constructing the representation of a read; (b) querying the vector store; (c) running fine-grained alignment with respect to the $K$ returned reference fragment sequences:

\begin{align*}
\mathcal{O}(F^2 + FQK + \epsilon(G,K)) \Rightarrow \mathcal{O}(F(F + QK) + \epsilon(G,K)) \\
\Rightarrow \mathcal{O}(FQK + \epsilon(G,K))
\end{align*}

% \subsection{Memory Cost}
% % guesstimate: will need to check 

% For the required setup we will need to embed the entirety of the reference genome assuming a given length of the reference samples $R$ with a reasonable overlap. We get that the memory cost os proportional to the size of the reference genome:

% \begin{equation}
%     \mathcal{O}(G)
% \end{equation}

% Typically proportional to some hyperparameters on the degree to which the database look-up should approximate a kNN.

\section{C. Ablation studies} \label{app:ablation}

\subsection{Without diversity priors in top-K} \label{app:diversity}
In the main text, we report the performance without diversity priors used in the retrieval step: i.e. the nearest-K neighbors in the embedding space are selected from the \textit{entire} set of fragments spanning the genome rather than uniformly sampling from each chromosome. The performance predicably falls in comparison to those reported in the main text since fewer fragments scattered unevenly across the different chromosomes are being retrieved per read.

\begin{table*}[ht]
\centering
\begin{tabular}{ll|ccc|ccc}
\toprule
 &  & \multicolumn{3}{c|}{$Q_{PH} \in [30, 60]$} & \multicolumn{3}{c}{$Q_{PH} \in [60, 90]$} \\
$I$ & $D$ & $d_{SW} < 1\%$ & $d_{SW} < 2\%$ & $d_{SW} < 5\%$ & $d_{SW} < 1\%$ & $d_{SW} < 2\%$ & $d_{SW} < 5\%$  \\
\midrule
\midrule
 \multicolumn{8}{c}{Top-1250 $\sim$ 50 x \{Chr. 22, X, Y, M\}} \\
\midrule
\multirow[t]{1}{*}{0.0} & 0.0 & {\cellcolor[RGB]{232,237,255}} 98.56 $\pm$ 0.67 & {\cellcolor[RGB]{196,215,255}} 97.60 $\pm$ 0.85 & {\cellcolor[RGB]{173,203,255}} 98.56 $\pm$ 0.67 & {\cellcolor[RGB]{214,228,255}} 97.76 $\pm$ 0.82 & {\cellcolor[RGB]{196,215,255}} 97.76 $\pm$ 0.82 & {\cellcolor[RGB]{177,205,255}} 97.44 $\pm$ 0.88  \\
\multirow[t]{1}{*}{0.01} & 0.01 & {\cellcolor[RGB]{232,237,255}} 98.8 $\pm$ 0.62 & {\cellcolor[RGB]{199,217,255}} 98.4 $\pm$ 0.71 & {\cellcolor[RGB]{179,207,255}} 98.8 $\pm$ 0.62 & {\cellcolor[RGB]{209,226,255}} 99.2 $\pm$ 0.52 & {\cellcolor[RGB]{189,211,255}} 98.4 $\pm$ 0.71 & {\cellcolor[RGB]{173,203,255}} 98.4 $\pm$ 0.71 \\

\midrule
 \multicolumn{8}{c}{Top-50 \textit{global}} \\
\midrule
\multirow[t]{1}{*}{0.0} & 0.0 & {\cellcolor[RGB]{232,237,255}} 91.6 $\pm$ 1.49 & {\cellcolor[RGB]{196,215,255}} 92.4 $\pm$ 1.43 & {\cellcolor[RGB]{173,203,255}} 93.8 $\pm$ 1.31 & {\cellcolor[RGB]{214,228,255}} 90.4 $\pm$ 1.58 & {\cellcolor[RGB]{196,215,255}} 93.8 $\pm$ 1.31 & {\cellcolor[RGB]{177,205,255}} 94.8 $\pm$ 1.21  \\
\multirow[t]{1}{*}{0.01} & 0.01 & {\cellcolor[RGB]{232,237,255}} 89.6 $\pm$ 1.64 & {\cellcolor[RGB]{199,217,255}} 91.6 $\pm$ 1.49 & {\cellcolor[RGB]{179,207,255}} 96 $\pm$ 1.07 & {\cellcolor[RGB]{209,226,255}} 94.4 $\pm$ 1.25 & {\cellcolor[RGB]{189,211,255}} 94.4 $\pm$ 1.25 & {\cellcolor[RGB]{173,203,255}} 93.2 $\pm$ 1.36 \\
\bottomrule
\end{tabular}
\vspace{1em}
\caption{\textbf{RDE performance on ESA - without diversity priors:} The various parameters are described
in the main text. It is evident that the addition of diversity priors results in an improvement of $\sim $ in recall. Similar to the result presented in the main text, performance improves with larger search radius in the vector store (top-$K$), higher quality reads ($Q_{PH}$)
and large distance bound ($d_{SW}$).}
\label{app:tab:diversity}
\end{table*}

\subsection{Across read lengths} \label{app:readlength}
In Table~\ref{app:tab:readlength}, the performance of RDE on ESA is reported across read lengths. We observe \textit{Zero-shot performance at longer read lengths}: The model performs better at longer read lengths (even exceeding the read length bound established during training $\mathcal{U}[150,500]$); while evaluating for longer reads, we make sure to guarantee that the reads exist as subsequences of fragments. Improving the performance at shorter read lengths is the subject of future work.

% Please add the following required packages to your document preamble:
% \usepackage{booktabs}
% \usepackage{multirow}
% \usepackage[table,xcdraw]{xcolor}
% If you use beamer only pass "xcolor=table" option, i.e. \documentclass[xcolor=table]{beamer}
\begin{table*}[t]
\centering

\begin{tabular}{@{}llccc@{}}
\toprule

\multicolumn{1}{l|}{I}    & \multicolumn{1}{l|}{D}    & \multicolumn{1}{c|}{$Q=200$}       & \multicolumn{1}{c|}{$Q=225$}       & \multicolumn{1}{c}{\textbf{$Q = 250$}}     \\ \midrule
\multicolumn{5}{c}{$Q_{PH} \in [30,60]$}                                                                                                                           \\ \midrule

\multicolumn{1}{l|}{0}    & \multicolumn{1}{l|}{0}    & \multicolumn{1}{c|}{99 $\pm$ 0.57} & \multicolumn{1}{c|}{98.6 $\pm$ 0.67} & \multicolumn{1}{c}{\textbf{99 $\pm$ 0.57}}                                                    \\ 
\multicolumn{1}{l|}{0.01} & \multicolumn{1}{l|}{0.01} & \multicolumn{1}{c|}{97.4 $\pm$ 0.88}  & \multicolumn{1}{c|}{98.8 $\pm$ 0.62} & \multicolumn{1}{c}{\textbf{98.6 $\pm$ 0.67}} 
\\ \midrule
\multicolumn{5}{c}{$Q_{PH} \in [60,90]$}                                                                                                                                 \\ \midrule
\multicolumn{1}{l|}{0}    & \multicolumn{1}{l|}{0}    & \multicolumn{1}{c|}{98.6 $\pm$ 0.67} & \multicolumn{1}{c|}{98.6 $\pm$ 0.67} & \multicolumn{1}{c}{\textbf{98.9 $\pm$ 0.59}}                                                    \\ 

\multicolumn{1}{l|}{0.01} & \multicolumn{1}{l|}{0.01} & \multicolumn{1}{c|}{98.6 $\pm$ 0.67} & \multicolumn{1}{c|}{99 $\pm$ 0.57} & \multicolumn{1}{c}{\textbf{98.6 $\pm$ 0.67}} 
\\ \bottomrule
\end{tabular}
\vspace{1em}
\caption{\textbf{RDE recall performance on ESA across read lengths:} Performance of RDE is robust across various read lengths supported by the ART simulator. Additional experiments performed using synthetically-generated ``pure'' reads -- random subsequences of the reference genome -- of length $[500,1000]$ showed that the recall performance is high ( $ \ge 99\%$). }
 
\label{app:tab:readlength}
\end{table*}

\section{D. Task transfer from Chromosome 2}

% Are these results indicative of DNA-ESA's adaptability to new genomic sequences rather than strict adherence to its training data? This would suggest the model's learning to solve the sequence alignment task rather than memorizing the genome.

We present two experiments that demonstrate that the model learns to solve the sequence alignment task rather than memorizing the genome on which it is trained.

\textbf{Experiment setup:} RDE is trained on Chromosome 2 -- the longest chromosome -- and recall is computed on unseen chromosomes from the human genome (3, Y) (\textit{inter-chromosome}) and select chromosomes from chimpanzee (2A,2B) and rat (1,2) DNA (\textit{inter-species}). Reads are generated with the following simulator configurations: $I=10^{-2}$, $D=10^{-2}$, $d_{SW}=2\%$, $Q=250$. Top-$K$ is set to $50$ / Chr., $\text{reads per setting}=5,000$. Independent vector stores are constructed for each chromosome; representations for reference fragments (staging) and reads (testing) \textit{are generated by the Chr. 2-trained model}. The results are reported in Tables~\ref{tab:chr2a}, \ref{tab:chr2b}.

% Please add the following required packages to your document preamble:
% \usepackage{multirow}

% Please add the following required packages to your document preamble:
% \usepackage{multirow}
% \begin{table}[h]
% \centering
% \begin{tabular}{l|c|c|c|c}
% \toprule
% Train                   & Reads & 2 & 3 & Y \\ \hline
% \multirow{2}{*}{Chr. 2} & Pure  & {\cellcolor[RGB]{232,237,255}} 97.5±0.46 & {\cellcolor[RGB]{214,228,255}} 98.2±0.41 & {\cellcolor[RGB]{173,203,255}} 99.3±0.27 \\ \cline{2-5} 
%                         & ART   & {\cellcolor[RGB]{220,230,255}} 97.9±0.43 & {\cellcolor[RGB]{227,234,255}} 97.6±0.44 & {\cellcolor[RGB]{214,228,255}} 98.0±0.43 \\ \bottomrule
% \end{tabular}
% \end{table}

\begin{table*}[t]
\setlength{\tabcolsep}{3pt} % Adjust the column separation here
\centering
\small % Reduce font size to small
\begin{tabular}{@{}l|cccc c@{}}
\toprule
\multicolumn{1}{c|}{}                                & \multicolumn{5}{c}{Vector Store}                                                              \\ \midrule
\multicolumn{1}{c|}{Read Origin}                   & \multicolumn{1}{c|}{Human Chr.2} & \multicolumn{1}{c|}{Human Chr.3} & \multicolumn{1}{c|}{Human Chr.Y} & \multicolumn{1}{c|}{Chimp Chr.2A} & Chimp Chr.2B \\ \midrule
\multicolumn{1}{l|}{Human Chr. 2} 
                      & \multicolumn{1}{c|}{\textbf{99.5 ± 0.4}} & \multicolumn{1}{c|}{1.5 ± 1.0} & \multicolumn{1}{c|}{$<$ 1} & \multicolumn{1}{c|}{34.5 ± 6.0} & 43.1 ± 7.0 \\ \midrule
\multicolumn{1}{l|}{Human Chr. 3} 
  & \multicolumn{1}{c|}{1.0 ± 0.8} & \multicolumn{1}{c|}{\textbf{99.0 ± 0.8}} & \multicolumn{1}{c|}{$<$ 1} & \multicolumn{1}{c|}{1.0 ± 0.8} & $<$ 1 \\ \midrule
\multicolumn{1}{l|}{Human Chr. Y} 
  & \multicolumn{1}{c|}{1.2 ± 0.9} & \multicolumn{1}{c|}{1.2 ± 1.1} & \multicolumn{1}{c|}{\textbf{98.75 ± 1.00}} & \multicolumn{1}{c|}{2.0 ± 1.9} & $<$ 1 \\ \midrule
\multicolumn{1}{l|}{Chimp Chr. 2A} 
  & \multicolumn{1}{c|}{70.0 ± 6.0} & \multicolumn{1}{c|}{$<$ 1} & \multicolumn{1}{c|}{$<$ 1} & \multicolumn{1}{c|}{\textbf{97.0 ± 1.9}} & 1.5 ± 1.1 \\ \midrule
\multicolumn{1}{l|}{Chimp Chr. 2B} 
  & \multicolumn{1}{c|}{71.0 ± 6.0} & \multicolumn{1}{c|}{$<$ 1} & \multicolumn{1}{c|}{$<$ 1} & \multicolumn{1}{c|}{1.0 ± 0.8} & \textbf{95.9 ± 2.0} \\ \bottomrule
\end{tabular}
\vspace{1em}

\caption{\small
\textbf{ESA Performance on Cross-Species Chromosome Alignment:} RDE model trained on human chromosome-2 is evaluated for aligning reads from human chromosomes 2, 3, Y, and chimpanzee chromosomes 2A and 2B to vector stores of each these chromosomes. Values represent the percentage of reads aligned to each store (mean ± standard deviation where applicable). The results show high accuracy for intra-species alignment (99.5\% for Human Chr. 2, 99.0\% for Human Chr. 3, 98.75\% for Human Chr. Y, 97.0\% for Chimp Chr. 2A, and 95.9\% for Chimp Chr. 2B). Notably, there's significant cross-species alignment between Human Chr. 2 and both Chimp Chr. 2A and 2B (34.5\%, 43.1\% respectively), and vice versa (70.0\%, 71.0\%), reflecting their evolutionary relationship. These findings indicate the RDE model's capability of aligning both positive and null reads.}
\label{tab:chr2a}
\end{table*}

\begin{table*}[t]
\setlength{\tabcolsep}{3pt} % Adjust the column separation here
\centering
\begin{tabular}{@{}l|c|c@{}}
\toprule
\textbf{Species}                        & $Q_{PH} \in [30, 60]$           & $Q_{PH} \in [60, 90]$            \\ \midrule
                     &  \multicolumn{2}{c}{Recall Top-K @50 $\uparrow$}            \\ \midrule
\rowcolor[HTML]{E8E9FE} 
\textit{Thermus Aquaticus {[}All{]}}    & 99.9 $\pm$ 0.01 & 99.9 $\pm$ 0.01  \\ \midrule
\rowcolor[HTML]{E8E9FE} 
\textit{Acidobacteriota {[}All{]}}      & 99.9 $\pm$ 0.01  & 99.9 $\pm$ 0.01  \\ \midrule
\textit{Rattus Norwegicus {[}Chr. 1{]}} & 95.98±1.07               & 96.48±1.01               \\ \midrule
\textit{Rattus Norwegicus {[}Chr. 2{]}} & 97.99±0.78               & 96.49±1.01               \\ \midrule
\rowcolor[HTML]{ECF4FF} 
\textit{Pan Troglodytes {[}Chr. 2A{]}}  & 98.5±0.69               & 95.74±1.11               \\ \midrule
\rowcolor[HTML]{ECF4FF} 
\textit{Pan Troglodytes {[}Chr. 2B{]}}  & 95.99±1.08               & 96.25±1.05               \\ \bottomrule
\end{tabular}
\vspace{1em}

\caption{\textbf{Cross-Species Task Transfer with RDE on ESA: Training on Human Chr. 2, Testing on Diverse Species:} RDE, trained on human Chr. 2, aligns fragments and reads from different species, including \textit{Rattus Norwegicus, Pan Troglodytes, Acidabacteriota}, and \textit{Thermus Aquaticus}. These species, are evaluated for read alignment recall using ART-generated reads. The findings indicate RDE's proficiency in modeling DNA sequence structure, beyond simply memorizing training data.}

\label{tab:chr2b}
\vspace{-1em}
\end{table*}

% \begin{table}[h]
% \setlength{\tabcolsep}{3pt} % Adjust the column separation here
% \centering
% \begin{tabular}{cc|ccc|cc|cc}
% \toprule
% \multicolumn{2}{c|}{} &
%   \multicolumn{3}{c|}{\textit{Homo Sapiens}} &
%   \multicolumn{2}{c|}{\textit{Pan Troglodyte}} &
%   \multicolumn{2}{c}{\textit{Rattus Norwegicus}} \\ \midrule
% \multicolumn{1}{c|}{Train} &
%   Reads &
%   \multicolumn{1}{c}{2} &
%   \multicolumn{1}{c}{3} &
%   \multicolumn{1}{c|}{Y} &
%   \multicolumn{1}{c}{2A} &
%   2B &
%   \multicolumn{1}{c}{1} &
%   2 \\ \midrule
% \multicolumn{1}{c|}{\multirow{2}{*}{Chr. 2}} &
%   Pure &
%   \multicolumn{1}{c}{\cellcolor[HTML]{C5E4C7} 97.5±0.46} &
%   \multicolumn{1}{c}{\cellcolor[HTML]{CDEECC} 98.2±0.41} &
%   \multicolumn{1}{c|}{\cellcolor[HTML]{E4F6E2} 99.3±0.27} &
%   \multicolumn{1}{c}{\cellcolor[HTML]{BFDDBE} 94.5±0.67} &
%   \cellcolor[HTML]{C3E1C6} 94.3±0.68 &
%   \multicolumn{1}{c}{\cellcolor[HTML]{B9D8B9} 93.3±0.73} &
%   \cellcolor[HTML]{C0DFC0} 93.2±0.74 \\ %\cline{2-10} 
% \multicolumn{1}{c|}{} &
%   ART &
%   \multicolumn{1}{c}{\cellcolor[HTML]{D7F3D5} 97.9±0.43} &
%   \multicolumn{1}{c}{\cellcolor[HTML]{D3F1D1} 97.6±0.44} &
%   \multicolumn{1}{c|}{\cellcolor[HTML]{D7F3D5} 98.0±0.43} &
%   \multicolumn{1}{c}{\cellcolor[HTML]{CFEDCF} 97.1±0.51} &
%   \cellcolor[HTML]{D1EFD1} 97.2±0.50 &
%   \multicolumn{1}{c}{\cellcolor[HTML]{C9E8C9} 95.4±0.62} &
%   \cellcolor[HTML]{CCEBCC} 96.4±0.55 \\ \bottomrule
% \end{tabular}
% \end{table}

\textbf{Performance:} Details on convergence are in the SI Sec. A. Table~\ref{tab:chr2a} shows that the performance on unseen human chromosomes (3, Y) is similar to the performance reported in the main table training RDE on Chr. 2. This suggests RDE's ability to generalize sequence alignment across chromosomes with different compositions. Even distantly related species like Thermus aquaticus and Acidobacteriota show significant recall, highlighting RDE's task transferability beyond simple data memorization. 

\section{E. Potential for speed-ups in index creation and read alignment}

The process of creating the ESA index involves generating embeddings for all reference fragments (Step 1), and uploading these embeddings to a vector database (Step 2). Traditional aligners like Bowtie-2 typically require 1–2 hours to construct an FM-index for the human genome on standard hardware. Currently, our RDE approach completes Step 1 in about 1 hour using a single NVIDIA A6000 GPU and takes an additional 1–2 hours to populate the Pinecone vector store. We are exploring several strategies to improve speed and efficiency without sacrificing performance:

\begin{itemize}
\item \textit{Parallelized Embedding Generation:} Leveraging multi-GPU setups and high-memory GPUs could parallelize Step 1, potentially achieving a 16x speedup with an 8-GPU cluster and 2x memory capacity in each GPU.

\item \textit{High-Throughput Vector Stores: }Utilizing vector databases with faster upload and polling rates would enable simultaneous read/write of multiple fragments/embeddings, albeit at higher cloud costs. Local vector stores like FAISS are another option.
\item \textit{Model Optimization:} Techniques such as model distillation and weights quantization of the embedding model could help accelerate embedding model inference.
\end{itemize}
\color{black}

\section{F. A setup for  de-novo Genome Assembly using RDE}

Consider a set of reads, $R := \{r_1, r_2, \dots, r_C\}$ from which we would like to estimate a reference $\mathcal{R}$. The de-novo assembly task is to estimate an ordering of the reads in $R$, $\hat{R} := (r_i, \dots, r_j, \dots, r_k)$ such that a concatenation (after removing overlaps) is approximately the original reference. In typical setups, all pairs of reads would need to be compared in order to find those that overlap and construct longer chains of the assembly in an incremental fashion. This may be visualized as a complete graph $K_C$ with $C$ nodes (reads) and $C^2$ edges (constraints). To avoid the quadratic complexity, modern techniques such as \textit{Overlap-Layout-Consensus} (OLC) or \textit{de Bruijn Graph} (DBG) approaches~\cite{assembly} have adopted hot-start heuristics such as base overlap and shared K-mer counts to reduce the number of pairwise read comparisons. 

The representation space modeled by RDE admits a comparable assembly setup. Fig. 2 suggested emergent 1D manifolds in the representation space -- comprising the embeddings of the reads -- $h(r_1), h(r_2), \dots, h(r_C)$ -- such that the relative \textit{positions} of fragments along the manifold correlated to the relative \textit{locations} of the fragments along the reference. \textit{Therefore, a pairwise read-read comparison within a smaller radius in the embedding space would be sufficient to find reads that overlap}. And more generally, a \textit{walk} along the 1D manifold would constitute a near-optimal assembly.

\begin{figure*}
    \centering
    \includegraphics[width=1.0\textwidth]{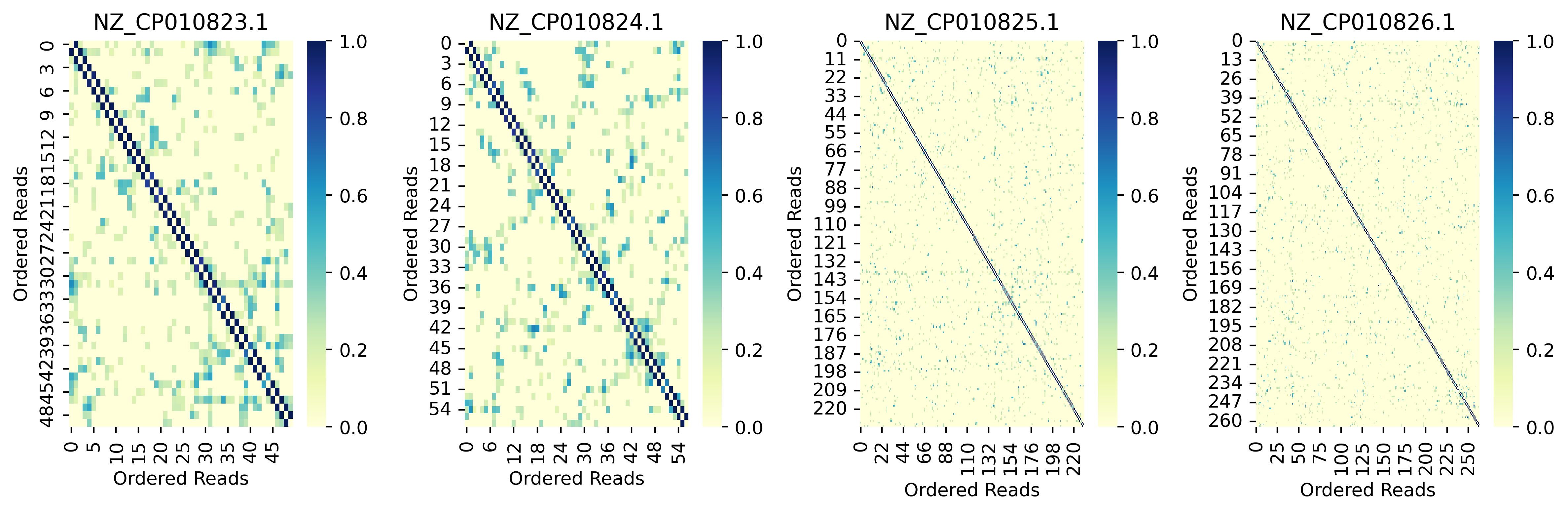}
    \caption{\textbf{Motivating the task of de-novo assembly using the distance adjacency matrix of the UMAP-generated $k$-nearest neighbor ($k=10$) network of reads from Thermus Aquaticus:} We assume that reads $r_1, r_2, \dots, r_C$ belonging to a species are embedded into the RDE embedding space $h(r_1), h(r_2), \dots, h(r_C)$ using a model trained on a known reference genome belonging to a different species.
    %for this example we use simulated reads from Thermus Aquaticus \cite{} and a model trained on the human genome as presented in Section. ~\ref{}. 
    For this illustrative numerical example we ensure that reads are created with the following properties: first, no read is a substring of any other read (efficient algorithms for the case where reads can be substrings of others will be covered in our future work); second every read has a corresponding read that it overlaps with for each end, and finally, the union of the reads cover the ground-truth reference genome. Thus, for every  read $r_i$ there exist reads $r_k$ and $r_j$ such that the ordered triplet $(r_j, r_i, r_k)$ constitutes a fragment of the reference. 
    %That is, the left end of $r_i$ ($r_k$) has partial overlap with the right end of $r_j$ ($r_i)$. 
    \textit{In a perfect embedding scenario}, the closest neighbors of $r_i$ (i.e. $k=1$ in an $k$-NN search) in the embedding space should be $r_j$ ad $r_k$.  For this ideal  case,  if a network is constructed where  each read (node) is connected to its two nearest neighbors, and the nodes are arranged based on the order in which they appear in the reference chromosome, one would get a bi-diagonal adjacency matrix. When  ground truth is not known,  the adjacency matrix will be a permuted version of a bi-diagonal matrix, but a complete assembly can still be performed by a greedy algorithm that starts with any node and pieces together the neighboring reads in the embedded space in time that grows linearly in the number of reads. Simulated reads, each  of length $500$ and with random overlaps of lengths drawn uniformly $\mathcal{U}([150,250])$ with its neighboring reads at each end, are sampled  from $4$ of the chromosomes in Thermus Aquaticus, and  are embedded into the RDE embedding space trained with a reference human genome. The sorted adjacency matrix of the UMAP~\cite{umap}-generated $k$-NN ($k=10$) network is presented above for each chromosome. The dark off-diagonal distance values in the $k$-NN adjacency graph with $k=10$ \textit{show that the distance properties of actual embeddings of simulated reads from Thermus Aquaticus are almost coincident with that of the ideal case discussed above}. Moreover, just as in the ideal case,  by finding an approximate solution to the Traveling Salesman Problem in the unsorted network, one can find a walk (the assembly) that is close to the original reference. The results are presented in Table~\ref{tab:assembly}.}
    \label{fig:chromosomefresh}
\end{figure*}

We provide an initial computational implementation of this idea: (a) The k-nearest neighbor (kNN) graph $G(R, d)$, representing the distances between close-by reads, is computed using the density-aware corrections provided by the UMAP library~\cite{umap}: graph adjacency matrices for each of the chromosomes is visualized in Fig.~\ref{fig:chromosomefresh}. For this figure, the several reads are ordered manually as they appear in the assembly in order to highlight the connectivity pattern; the bright off-diagonal values indicate a chain-like structure corresponding to the observed manifold; (b) A \textit{Hamiltonian} cycle computed through the graph that has the lowest total distance corresponds to an assembly, $\hat{R}$. This involves solving the Traveling Salesman Problem (TSP) -- we use the 3/2-approximate Christofides algorithm~\cite{christofides}. After a walk across the reads is generated, a second pass through the walk greedily concatenates adjacent reads $(r_i, r_{i+1})$ by computing the alignment using the SW distance and removing duplicates and overlaps. In cases where adjacent reads do not have any alignment to one another -- the walk isn't guaranteed to be perfect since the RDE model is heuristical and the TSP algorithm is approximate -- the current assembly part is stowed away as a \textit{contig}~\cite{contig}, and a new contig is instantiated with the incoming read. The fewer the contigs, the better the manifold in the representation space correlates to the ordering of the true reference.

Assembly is conducted across $4$ of the shorter chromosomes in the \textit{Thermus Aquaticus} genome. We consider (long) reads of length $Q=500$, which have no indels, and the entire corpus of reads is ensured to have full coverage across each chromosome. Each read has an overlap of $\mathcal{U}([150,250])$ bases with another read in the corpus of reads. QUAST~\cite{quast} metrics are reported in Table~\ref{tab:assembly}. 

% Please add the following required packages to your document preamble:
% \usepackage{booktabs}
% \usepackage{multirow}
\begin{table*}[h]
\centering
\begin{tabular}{@{}cl|c|c|c|c@{}}
\toprule
\multicolumn{2}{l|}{} & \textit{\textbf{Size (in kb) $\uparrow$}} & \textit{\textbf{No. of Contigs} $\downarrow$} & \textit{\textbf{N50 (in kb)} $\uparrow$} & \textit{\textbf{NGA50 (in kb)} $\uparrow$} \\ \midrule
\multicolumn{1}{c|}{\multirow{4}{*}{\textbf{\begin{tabular}[c]{@{}c@{}}Thermus \\ Aquaticus\end{tabular}}}} & NZ\_CP010823.1 & 14.4 & 1 & 14.4 & 14.4 \\ \cmidrule(l){2-6} 
\multicolumn{1}{c|}{} & NZ\_CP010824.1 & 16.6 & 1 & 16.6 & 16.6 \\ \cmidrule(l){2-6} 
\multicolumn{1}{c|}{} & NZ\_CP010825.1 & 69.9 & 22 & 6.0 & 4.4 \\ \cmidrule(l){2-6} 
\multicolumn{1}{c|}{} & NZ\_CP010826.1 & 78.7 & 15 & 10.3 & 10.3 \\ \bottomrule
\end{tabular}
\vspace{1em}

\caption{\textbf{RDE performance on de-novo assembly of the Thermus Aquaticus genome:} The solution to the approximate TSP applied on the RDE read-read network generates a set of contigs that can be evaluated with respect to the original reference using QUAST~\cite{quast}. For the two shorter chrosomomes, the resulting assembly was perfect; the entire reference is encapsulated by one contig. Even in the case with longer chromosomes, the number of contigs are still few and of significant length (as indicated by the high N(G)50 score) suggesting a strong correlation between the distance metric in the embedding space and the ordering of the corresponding reads in the assembly.}
\label{tab:assembly}
\end{table*}

N50 refers to the length of a contig (in bases) such that the contigs that exceed this length span 50\% of the reference genome. NGA50 is a reference-aware N50 metric. The fewer contigs, large N(GA)50 scores in comparison to the large size of the chromosomes, when considered together, demonstrate that the reads are indeed ordered in a relative position-aware manner in the RDE embedding space with respect to a reference making the de-novo assembly task viable.

\end{appendices}

\end{document}